\documentclass[amsmath,amssymb,twocolumn]{aastex63}

\usepackage{textcomp}
\usepackage{gensymb}
\usepackage[hang,flushmargin,stable]{footmisc} 
\usepackage{graphicx}
\usepackage{txfonts}
\usepackage{xcolor}
\usepackage{float}
\usepackage{hyperref}
\usepackage{bm}
\usepackage{tabularx}
\usepackage{multirow}
\newcommand{\orcid}[1]{\href{https://orcid.org/#1}{\includegraphics[width=8pt]{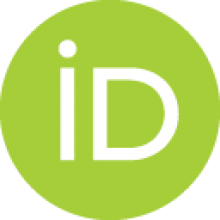}}}

\shorttitle{CV Evolution with Binary Driven Mass-Loss}
\shortauthors{Tang, Li, \& Cui}
\graphicspath{{./}{Figures}}
\begin{document}

\title{Evolution of Cataclysmic Variables with Binary-Driven Mass-Loss during Nova Eruptions}

\author{Wen-Shi Tang\orcid{0000-0002-6588-9264}}
\affiliation{School of Astronomy and Space Science, Nanjing University, Nanjing 210023, China; \href{mailto:tangwenshi20@163.com}{tangwenshi20@163.com},\href{mailto:lixd@nju.edu.cn}{lixd@nju.edu.cn}}
\affiliation{Key Laboratory of Modern Astronomy and Astrophysics (Nanjing University), Ministry of Education, Nanjing 210023, China}
\affiliation{Department of Astronomy, Xiamen University, Xiamen 361005, China}

\author{Xiang-Dong Li\orcid{0000-0002-0584-8145}}
\affiliation{School of Astronomy and Space Science, Nanjing University, Nanjing 210023, China; \href{mailto:tangwenshi20@163.com}{tangwenshi20@163.com},\href{mailto:lixd@nju.edu.cn}{lixd@nju.edu.cn}}
\affiliation{Key Laboratory of Modern Astronomy and Astrophysics (Nanjing University), Ministry of Education, Nanjing 210023, China}

\author{Zhe Cui\orcid{0000-0001-8311-0608}}
\affiliation{School of Astronomy and Space Science, Nanjing University, Nanjing 210023, China; \href{mailto:tangwenshi20@163.com}{tangwenshi20@163.com},\href{mailto:lixd@nju.edu.cn}{lixd@nju.edu.cn}}
\affiliation{Key Laboratory of Modern Astronomy and Astrophysics (Nanjing University), Ministry of Education, Nanjing 210023, China}
\affiliation{College of Physics and Electronic Information, Dezhou University, Dezhou 253023, China}

\begin{abstract}
The discrepancies between observations and theoretical predictions of cataclysmic variables (CVs) suggest that there exists unknown angular momentum loss mechanism(s) besides magnetic braking and gravitational radiation. Mass loss due to nova eruptions belongs to the most likely candidates. While standard theory assumes that mass is lost in the form of radiation driven, optically thick wind (fast wind; FW), recent numerical simulations indicate that most of the mass loss is initiated and shaped by binary interaction. We explore the effect of this binary-driven mass-loss (BDML) on the CV evolutions assuming a major fraction of the lost mass leaves the system from the outer Lagrangian point. Different from the traditional continuous wind picture, we consider the mass loss process to be instantaneous, because the duration of nova eruptions is much shorter than the binary evolutionary timescale. Our detailed binary evolution calculations reveal the following results. (1) BDML seems able to provide extra angular momentum loss below the period gap. The mass transfer rates at a given orbital period occupy a large range, in agreement with the observed secular mass transfer rate distribution in CVs. (2) The enhanced mass transfer rates do not lead to runaway mass transfer process, and allow the white dwarfs to grow mass $\lesssim 0.1\,M_{\sun}$. (3) BDML can cause both positive and negative variations of the orbital period induced by nova eruptions, in line with observations, and can potentially explain the properties of some peculiar supersoft X-ray sources likely CAL 87, 1E 0035.4$-$7230, and RX J0537.7$-$7034.

\end{abstract}


\section{Introduction}\label{Sect_Intro}

Cataclysmic variables (CVs) are short-period binaries in which a white dwarf (WD) accretes mass from a low-mass main sequence (MS) companion star, usually of K-M spectral type \citep{2022abn..book.....C}. With typical accretion rates $\sim 10^{-10}-10^{-8}\,M_{\rm \sun}\,\rm yr^{-1}$, the accreted hydrogen-rich material cannot burn stably but accumulate on the surface of the WD \citep{2013ApJ...777..136W}. The increase in the mass of the hydrogen layer causes the density and temperature at its base to rise. After a critical mass is reached, the layer  undergoes unstable (runaway) nuclear burning which results in a drastic increase in luminosity. This is called nova eruption \citep[see][for a detailed review]{2021ARA&A..59..391C}. Some of them  were observed and recorded as ``guest stars'' in ancient China \citep[e.g.][]{1976QJRAS..17..121S, 2020MNRAS.494.5775H}. 
 
The evolution of CVs is mainly driven  by angular momentum loss and briefly summarized as follows \citep[see][for a comprehensive review]{1995cvs..book.....W}.  After emerging from the common envelope phase, the detached system composes of a WD and a low-mass main sequence star in an orbit of about a few days or less \citep[e.g., ][]{2007ApJ...665..663P}. Subsequently, the companion star starts to fill out its Roche lobe and transfer mass to the WD. Mass transfer is driven either by angular momentum loss that reduces the binary separation or expansion of the companion star, depending on the initial orbital period and the companion mass. If the companion star is initially more massive, mass transfer proceeds on a (sub)thermal timescale and the binary looks like a supersoft X-ray source \citep{KH97}. After the mass ratio reverses, the mass transfer rate declines, and the binary behaves as a CV. For the orbital periods around $3-10$\,hr, the dominant mechanism of angular momentum loss in CVs is magnetic braking \citep[MB; ][]{1972ApJ...171..565S, 1983ApJ...275..713R}. 
Because of unstable hydrogen nuclear burning, the WD experiences repeating nova eruptions, during which most or all of the accreted material is ejected from the binary. When the companion star reduces its mass to $\sim 0.2-0.3\,M_{\rm \odot}$, it becomes fully convective \citep[e.g.,][]{1972ApJ...171..565S, 1983ApJ...275..713R, 2011ASPC..447....3K}. The traditional postulation is that MB ceases at this point \citep[although there are pieces of evidence that fully convective stars can indeed possess strong magnetic fields, e.g.,][]{2000ApJ...538L.141R, 2020AdSpR..66.1226B}. The abrupt stop of MB makes the donor to shrink and become detached from its Roche lobe. The binary enters the phase of so-called ``period gap" (with orbital periods $\sim 2-3$\,hr), in which observations show a statistically significant deficit of CVs. In the period gap, the only mechanism of angular momentum loss is gravitational radiation (GR). The companion  fills out its Roche-lobe at the orbital period $\sim 2$ hr and initiates mass transfer once more. The orbital period continues decreasing during the mass transfer until a minimum value is reached. After that, the mass-loss timescale  of the donor becomes shorter than its thermal timescale and the donor stops shrinking in response to mass-loss. Consequently, the binary evolves back towards longer orbital periods, i.e., ``period bouncer". The above pattern depicts a standard pathway of CVs.
  
However, The standard theory faces severe difficulties in addressing several observational issues \citep[see][for a recent review]{2020AdSpR..66.1080Z}. While the standard theory predicts the minimum orbital period $P_{\rm orb,min}$ of CVs to be $\sim 65-70$\,min \citep{1999MNRAS.309.1034K}, observations indicate   $P_{\rm orb,min}\sim 82$\,min \citep{2009MNRAS.397.2170G,2011ApJS..194...28K,2019MNRAS.486.5535M}. Moreover, the standard theory predicts that $99\%$ of CVs should be below the period gap and over half of CVs have brown dwarf companions \citep{2001ApJ...550..897H, 2015ApJ...809...80G}, while observations  indicate that the fractions of CVs below and above the orbital period gap are 83$\pm$6\% and 17$\pm$6\%, respectively \citep{2003A&A...404..301R,2020MNRAS.494.3799P}. Isolated WDs and WDs in post-common envelope binaries which are the progenitors of CVs, both have a mean mass $\sim 0.5\,M_{\rm \odot}$, consistent with theoretical predication \citep{1992A&A...261..188D,1993A&A...267..397D}, but the mean mass of WDs in CVs is significantly larger, $\sim 0.83\,M_{\rm \odot}$ \citep{2011A&A...536A..42Z, 2022RAA....22d5003Y}. 

The resolution to these discrepancies is likely related to the angular momentum loss mechanisms. \cite{2011ApJS..194...28K} pointed out that if the angular momentum loss rate below the period gap exceeds that solely caused by GR by a factor of 1.47, there would be better agreement between the observational and theoretical donor mass-radius relations. Including extra angular momentum loss could also reproduce the observed $P_{\rm orb, min}$. More recently, \cite{2016MNRAS.455L..16S} suggested an empirical form of enhanced consequential angular momentum loss (eCAML) form due to mass loss, with its rate inversely proportional to the WD mass. With the help of eCAML, they showed that CVs with lower mass WDs are more likely to experience unstable mass transfer, resulting in merger of the binary. This may simultaneously account for the above-mentioned discrepancies. However, it is unclear about the the  underlying physics of the empirical eCAML prescription. The possible mechanisms include friction of the nova envelope with the binary \citep{1998MNRAS.297..633S}, mass-loss from the outer Lagrange point $L_{2}$, or formation of a circumbinary disk \citep{2012ApJ...745..165S}. The envelope friction model 
may not be able to explain the extra angular momentum loss beyond GR by a factor of 1.47 below the period gap unless the expanding velocities  of the nova envelope are extremely low \citep{2019ApJ...870...22L}, and the nova outflow seems to be the more plausible origin of eCAML \citep{2016MNRAS.455L..16S}.
 
Recently, \cite{2022ApJ...938...31S} performed hydrodynamical simulations of classical nova outflow to investigate whether optically thick winds can be efficiently accelerated during nova eruptions. Their results show that, although a high-mass ($\ge 0.8\,M_{\rm \sun}$) WD can successfully launch radiation-pressure driven, optically thick winds,  most of these winds are generally accelerated at radii beyond the WD’s Roche-lobe radius. Moreover, lower-mass ($0.6\,M_{\rm \sun}$) WDs do not develop any unbound optically thick winds. This implies that the companion star plays a crucial role in driving mass-loss during nova eruptions, named as binary-driven mass-loss (BDML). The slow winds accelerated by the companion carry more angular momentum than the traditional fast winds (FW), and may have a profound effect on the evolution of CVs.  
In this work, we explore the impact of BDML during nova eruptions on the evolution of CVs, to examine whether it can help resolve the discrepancies between theory and observation.

The remainder of this paper is organized as follows. In Sect.\,\ref{Sect_method}, we present the assumptions and method of calculating angular momentum loss during nova eruptions. In Sect.\,\ref{Sect-results}, we demonstrate our numerically calculated results about the effect of mass loss on the CV evolution and compare them with observations. In Sect.\,\ref{Discussion}, we compare our results with previous theoretical works and discuss their observational implications. Finally, our summary is given in Sect.\,\ref{Sect-Con-Diss}.
 
\section{Assumptions and ethods}\label{Sect_method}

\subsection{Mass transfer and thermonuclear burning}
Mass transfer in CVs is driven by angular momentum loss due to GR, MB and mass loss. While the angular momentum loss rate via GR \citep{1959flme.book.....L} has been verified with high precision \citep{2005ASPC..328...25W,2006Sci...314...97K}, there are substantial uncertainties in the latter two. For MB, the traditional Skumanich law \citep{1972ApJ...171..565S, 1983ApJ...275..713R} is adopted in our calculation, although various forms of the MB laws have been proposed in the literature \citep[e.g.,][]{2012ApJ...754L..26M, 2012ApJ...746...43R, 2019ApJ...886L..31V,  2019MNRAS.483.5595V}. Mass loss-associated angular momentum loss will be discussed in Sect.\,\ref{Sect_AML-ML}. 

We adopt the \citet{1988A&A...202...93R} scheme to calculate the mass transfer rate via Roche lobe overflow. As the accreted hydrogen-rich gas accumulates on the surface of the WD, thermonuclear burning will ensue stably or unstably, dependent on the WD mass ($M_{\rm WD}$) and the mass transfer rate ($\dot M_{\rm 2}$) from its companion \citep[of mass ${M}_2$, see][for reviews]{1982ApJ...253..798N,2007ApJ...663.1269N, 2021ARA&A..59..391C}. There exist two threshold mass transfer rates
\begin{equation}\label{Eq-dotMstable}
 {\rm log} \dot M_{\rm stable} =1.471m_{\rm WD}^{3}-5.224m_{\rm WD}^{2}+7.039m_{\rm WD}-10,
 \end{equation}
 and 
 \begin{equation}\label{Eq-dotMcrit}
{\log } \dot M_{\rm crit} = 1.828m_{\rm WD}^{3}-6.145m_{\rm WD}^{2}+7.431m_{\rm WD}-9.483,
 \end{equation}
between which the transferred hydrogen-rich material can burn stably on the surface of the WD at a rate of $|\dot M_{2}|$. In Eqs.~(1) and (2), $\dot M_{\rm stable}$ and $\dot M_{\rm crit}$  are in the units of $M_{\rm \sun}\,\rm yr ^{-1}$, $m_{\rm WD}=M_{\rm WD}/M_{\rm \sun}$, and the coefficients are obtained by fitting the numerical results of \cite{2013ApJ...778L..32M}. 
During stable burning stage the WD can accumulate mass at a rate of 
 \begin{equation}
 \dot M_{\rm WD} = \eta_{\rm H}\eta_{\rm He}|\dot M_{2}|,
 \end{equation}
where $\eta_{\rm H}$ and $\eta_{\rm He}$ are the mass accumulation efficiencies for hydrogen and helium burning, respectively. We take the same form of $\eta_{\rm H}$ in \cite{2010MNRAS.401.2729W} and \cite{ 2019A&A...622A..35L}, 
 \begin{equation}
\eta _{\rm H}=\left\{
 \begin{array}{ll}
 \dot{M}_{\rm crit}/|\dot{M}_{\rm 2}|, & {\rm if}\ |\dot{M}_{\rm 2}|> \dot{M}_{\rm crit},\\
 1, & {\rm if}\ \dot{M}_{\rm crit}\geq |\dot{M}_{\rm 2}|\geq \dot{M}_{\rm stable},\\
\end{array}\right.
\end{equation}
and the calculated value of $\eta_{\rm He}$ by \cite{2017A&A...604A..31W}. Note that $\eta_{\rm He}$ is available only for $M_{\rm WD} \ge 0.7M_{\rm \sun}$ in \cite{2017A&A...604A..31W}. For less massive WDs ($M_{\rm WD}< 0.7\,M_{\rm \sun}$), we simply assume $\eta_{\rm He}=1$ \citep{2004ApJ...613L.129K, 2019ApJ...870...22L}.

If $|\dot M_{2}|>\dot M_{\rm crit}$, the burning rate cannot match the mass transfer rate and is limited to $\dot M_{\rm crit}$. The excess material will pile up into an extended red giant-like envelope or be lost in a radiation pressure driven wind.

If $|\dot M_{\rm 2}|<\dot M_{\rm stable}$, the accreted material cannot burn stably but accumulate on the surface of the WD. A nova eruption is triggered when a critical amount of mass ($\Delta M_{\rm ign}$) is reached. We assume that the material accumulated on the WD is completely ejected during nova eruptions, which means that there is no net WD mass growth in CVs.
 
After the nova eruption, the WD continues to accrete until the next nova eruption occurs. During the repeating cycles, the recurrence timescale ($\tau_{\rm rec}$) or the duration required to build up mass through accretion up to  $\Delta M_{\rm ign}$, depends on $M_{\rm WD}$ and $\dot M_{\rm 2}$. We fit the $\dot M-m_{\rm WD}$ relation for a given $\tau_{\rm rec}$ from the numerically calculated data by \citet[][Fig.~2 therein]{2021ARA&A..59..391C},

 \begin{equation}\label{Eq-t-rec}
\log \dot M=\left\{
 \begin{array}{ll}
 -1.3997m_{\rm WD}-7.0996, {\rm for}\,\ \tau_{\rm rec}=10^{4}\,{\rm yr}, \\
 -1.5016m_{\rm WD}-7.7514, {\rm for}\,\ \tau_{\rm rec}=10^{5}\,{\rm yr}, \\
 -1.6387m_{\rm WD}-8.5280, {\rm for}\,\ \tau_{\rm rec}=10^{6}\,{\rm yr}, \\
 -1.3463m_{\rm WD}-9.7792, {\rm for}\,\ \tau_{\rm rec}=10^{7}\,{\rm yr},\\
\end{array}\right.
\end{equation}

\noindent where the accretion rate $\dot M$ is in the units of $M_{\rm \sun}\,\rm yr ^{-1}$.  For $\tau_{\rm rec}\le10^{3}\,{\rm yr}$, we directly extract their data. Since the calculations by \citet{2021ARA&A..59..391C} are limited to the WD masses $\geq 0.6\,M_{\rm \odot}$, we also extrapolate the fitting results to WDs of mass $0.4\,M_{\rm \odot}$. Then, we perform 2D interpolation to obtain $\tau_{\rm rec}$ as a function of $M_{\rm WD}$ and $\dot M$ in the $\dot M_{2}-m_{\rm WD}$ plane, and $\Delta M_{\rm ign}=\dot M\times \tau_{\rm rec}$. Because the typical mass ejection timescale during novae eruptions is around weeks to years, much shorter than $\tau_{\rm rec}$ \citep{2021ARA&A..59..391C}, we assume that the material is instantaneously ejected from the system.

\subsection{Angular moment loss due to mass-loss}\label{Sect_AML-ML}
 
Mass loss during nova eruptions may proceed as fast wind from the WD, accelerated through the iron opacity bump \citep{1994ApJ...437..802K}. Angular momentum loss during nova eruptions consequently causes a change in the binary separation $a$ \citep[e.g.,][]{1986ApJ...311..163S}. The orbital angular momentum  of a binary is 
 \begin{equation}\label{Eq-Jorb}
 J_{\rm orb} = M_{\rm WD}M_{2}\sqrt{\frac{G a}{M}}.
 \end{equation}
 Differentiation of Eq.\,(\ref{Eq-Jorb}) gives
 \begin{equation}\label{Eq-dota}
 \frac{{\rm d}a}{a}=2\frac{{\rm d}J_{\rm orb}}{J_{\rm orb}}-2\frac{{\rm d}M_{\rm WD}}{M_{\rm WD}}-2\frac{{\rm d}M_{\rm 2}}{M_{\rm 2}}+\frac{{\rm d}M_{\rm WD}+{\rm d}M_{2}}{M},
 \end{equation}
 where $M=M_{\rm WD}+M_{2}$, and $G$ is the gravitational constant. Define the mass ratio $q=M_{2}/M_{\rm WD}$, Eq.\,(\ref{Eq-dota}) can be transformed to be
 \begin{equation}\label{Eq-dota-q}
 \frac{{\rm d}a}{a}=2\frac{{\rm d}J_{\rm orb}}{J_{\rm orb}}-\left(\frac{1+2q}{1+q}\right)\frac{{\rm d}M_{\rm WD}}{M_{\rm WD}}-\left(\frac{2+q}{1+q}\right)\frac{{\rm d}M_{\rm 2}}{M_{\rm 2}}.
 \end{equation}
During nova eruptions, material of mass $\Delta M_{\rm ign}$ is ejected from the WD,  i.e., ${\rm d}M_{\rm WD}=-\Delta M_{\rm ign}$. In this FW model, we neglect any re-capture of the nova ejecta by the companion star because of the high speed (up to 1000\,km\,s$^{-1}$) of the wind \citep{2008clno.book.....B}, i.e., ${\rm  d}M_{2}=0$. Therefore, we have 
\begin{equation}\label{Eq-delta-q}
 \frac{\Delta a}{a}=2\frac{\Delta J_{\rm orb}}{J_{\rm orb}}+\left(\frac{1+2q}{1+q}\right)\frac{\Delta M_{\rm ign}}{M_{\rm WD}}.
 \end{equation}
In principle, $\Delta J_{\rm orb}=\Delta J_{\rm GR}+\Delta J_{\rm MB}+\Delta J_{\rm ML}$.  However, the former two terms can be ignored because the nova eruption is instantaneous. Hence $\Delta J_{\rm orb}=\Delta J_{\rm ML}$, or, 
 \begin{equation}
    \Delta J_{\rm orb,FW}=-\left(\frac{M_{2}J_{\rm orb}}{MM_{\rm WD}}\right)\Delta M_{\rm ign}.
 \end{equation} 
 Then, 
\begin{equation}\label{Eq-dota-FW}
\frac{\Delta a}{a} = \left(\frac{\Delta M_{\rm ign}}{M_{\rm WD}}\right)\left(\frac{1}{1+q}\right).
\end{equation}
Eq.\,(\ref{Eq-dota-FW}) indicates that $\Delta a$ is always positive, that is, nova eruption will cause expansion of the binary separation and the companion’s Roche lobe. Thus, the companion will be detached from its Roche lobe, and the binary will enter a state with very low mass transfer rate, making it hardly detected. This is the so-called hibernation theory \citep{1986ApJ...311..163S}.

\cite{2020MNRAS.492.3343S} reported orbital period changes across nova eruptions in six classical nova systems. He found that in five of the six systems  the binary separations decreased across the nova eruptions. This is in direct contrast with the prediction by Eq.\,(\ref{Eq-dota-FW}) \citep[see also][]{2023MNRAS.tmp.2146S}. In addition, the observed properties of classical novae, including aspherical ejecta,  multiple modes of mass ejection,  and  the super-Eddington luminosities \citep{2014Natur.514..339C, 2018MNRAS.474.2679A, 2020ApJ...905...62A,2001MNRAS.320..103S}, also disagree with the prediction of the FW model \citep{2021ARA&A..59..391C}.

According to \cite{2022ApJ...938...31S}, in most novae the outflowing material gradually expands into the circumbinary environment and makes its way out of the outer Lagrange point $L_{2}$.  
To calculate the amount of angular momentum loss associated with mass loss from the $L_{2}$ point, we need to find the distance $a_{L_{2}}$ between the centre of mass of the binary and the $L_{2}$ point. We first utilize the formula provided by \cite{2020A&A...642A.174M} to obtain the distance $D_{L_{2}}$ between the centre of the companion and the $L_{2}$ point as follows
\begin{equation}
\frac{D_{L_{2}}(q>1)}{R_{\rm RL,2}} = 3.334q^{-0.514}e^{-0.052/q}+1.308,
\end{equation}
 and
\begin{equation}\
\frac{D_{L_{2}}(q\le 1)}{R_{\rm RL,2}} = -0.04q^{-0.866}e^{-0.040/q}+1.883,
\end{equation}
where $R_{\rm RL,2}$ is the Roche lobe radius of the companion. Then, we have
\begin{equation}\label{aL2-q-gt-1}
a_{L_{2}}(q>1) = D_{L_{2}}(q>1)-\frac{aq}{1+q},
\end{equation}
and
\begin{equation}\label{aL2-q-le-1}
a_{L_{2}}(q\le 1) = D_{L_{2}}(q<1)+\frac{a}{1+q}.
\end{equation}
The amount of angular momentum loss during a nova eruption is $\Delta J_{{\rm orb},L_{2}}=-\Delta M a_{L_{2}}^{2}\omega\cdot$, where $\omega$ is the orbital angular velocity of the binary and $\Delta M$ is the lost mass from the $L_2$ point. From Eq.\,(\ref{Eq-dota-q}), the associated variation in the binary separation is
 \begin{equation}\label{Eq-dota-L2-1}
 \frac{{\rm d}a}{a}=-\frac{2a_{L_{2}}^{2}\omega \Delta M}{J_{\rm orb}}-\left(\frac{1+2q}{1+q}\right)\frac{{\rm d}M_{\rm WD}}{M_{\rm WD}}-\left(\frac{2+q}{1+q}\right)\frac{{\rm d}M_{\rm 2}}{M_{\rm 2}}.
 \end{equation}
In the case of BDML, the wind material has a relatively low velocity, so part of the nova ejecta could be captured by the companion. We let  $\gamma_{\rm re}$ to be the fraction of the nova ejecta re-accreted by the companion, i.e., $\gamma_{\rm re}=1-\Delta M/\Delta M_{\rm ign}$. Then, Eq.\,(\ref{Eq-dota-L2-1}) can be written as
\begin{eqnarray}\label{Eq-dota-L2-2}
 \frac{{\Delta}a}{a} &= &-\frac{2a_{L2}^{2}\omega (1-\gamma_{\rm re})\Delta M_{\rm ign}}{J_{\rm orb}}
  +\left(\frac{1+2q}{1+q}\right)\frac{\Delta M_{\rm ign}}{M_{\rm WD}}
 \nonumber \\ 
 & & -\left(\frac{2+q}{1+q}\right)\frac{\gamma_{\rm re}\Delta M_{\rm ign}}{M_{\rm 2}}. 
\end{eqnarray}
We see from Eq.\,(\ref{Eq-dota-L2-2}) that, unlike in the hibernation theory, $\Delta a$ can be both positive and negative in this case.
 
The change in the orbital separation due to nova eruptions can also cause the variation of the companion's Roche-lobe radius. Using the expression for the equivalent Roche-lobe radius given  by \cite{1983ApJ...268..368E}, 
\begin{equation}\label{Eq-RL}
\frac{R_{\rm L}}{a}=\frac{0.49q^{2/3}}{0.6q^{2/3}+{\rm ln}(1+q^{1/3})},
\end{equation}
we obtain 
 \begin{equation}\label{Eq-dRL}
\frac{\Delta R_{\rm L}}{R_{L}} = \frac{\Delta a}{a}+\frac{2}{3}\frac{\Delta q}{q}-\frac{(\frac{2}{5}q^{-1/3}+\frac{1}{3}\frac{q^{-2/3}}{1+q^{1/3}})}{0.6q^{2/3}+{\rm ln}(1+q^{1/3})}\Delta q,
\end{equation}
with
\begin{equation}
\Delta q=\frac{\Delta M_{\rm ign}}{M_{\rm WD}}(\gamma_{\rm re}+q).
\end{equation}
For simplicity, we assume $\gamma_{\rm re}=0$ in the following calculations and discuss its possible influence in section \ref{Sect-re-acc}.

Calculations by \cite{2022ApJ...938...31S} show that, for a $0.6\,M_{\rm \sun}$ WD, all of the accreted material is ejected through binary interaction, while for a WD with mass $\ge 0.8\,M_{\rm \sun}$, although most of mass loss is driven by binary interaction, there are still fast winds being successfully launched. We therefore construct a hybrid FW+BDML model, and define $f_{\rm ML, L2}$ and $f_{\rm ML, FW}$ ($= 1-f_{\rm ML, L2}$) to be the fraction of the ejected mass through BDML and fast wind, respectively. The change in the orbital angular momentum in a nova eruption is then
\begin{equation}\label{DJ-FW-L2}
\Delta J_{\rm orb} = f_{\rm ML, FW} \times \Delta J_{\rm orb, FW}+f_{\rm ML, L2}\times \Delta J_{\rm orb, L2}.
\end{equation}
We extract the data in Fig.\,8 of \cite{2022ApJ...938...31S} for $M_{\rm WD}=0.6,0.8,1.0,1.2\,M_{\rm \odot}$ to get $f_{\rm ML, L2}$ as a function of $\Delta M_{\rm ign}$ by interpolation, and then extrapolate to different WD mass ranges (see Appendix \ref{Appenx-f-L2}).
Combining Eqs.\,(\ref{Eq-dota-L2-2}),  (\ref{Eq-dRL}), and (\ref{DJ-FW-L2}), we obtain the net variations in both the binary separation and the Roche-lobe radius of the companion star for each cycle of nova eruption.

We emphasize that the above considerations are limited to the regime of nova eruptions (i.e., $|\dot M_{2}|<\dot M_{\rm sable}$).
If $|\dot M_{2}|>\dot M_{\rm crit}$, 
we still let the excess material (at a rate of $|\dot M_{2}| - \dot M_{\rm crit}$) leave the binary in the form of continuous fast winds. 
 
\section{Results}\label{Sect-results}

We first use the stellar evolution code {\tt MESA} (version 11701; \citealt{2011ApJS..192....3P, 2013ApJS..208....4P,2015ApJS..220...15P,2019ApJS..243...10P}) to follow the detailed evolution of WD+MS binaries. We take the initial WD masses $M_{\rm WD,i}$ to be 0.4, 0.6, 0.8, 1.0, and 1.2\,$M_{\rm \sun}$, and the donor stars to have solar metallicities. The results are then combined with binary population synthesis to investigate the statistical characteristics of CVs.

\subsection{The detailed binary evolution results}

To examine the influence of nova eruptions on the evolution of CVs, the timestep in the calculation should be short enough ($<\tau_{\rm rec}$), so that they can be resolved. In {\tt MESA}, the timestep is limited to {\it max\_timestep}, which is adjusted automatically if  {\it max\_timestep} $\le 0$. We let the timestep control to be
 \begin{equation}\label{Eq-basic-dt}
max\_timestep =\left\{
 \begin{array}{ll}
  f_{\rm dt}\tau_{\rm rec}, & {\rm if}~|\dot M_{2}|<\dot M_{\rm stable}\\
 0, & {\rm if}~|\dot M_{2}|\ge \dot M_{\rm stable}~{\rm or}~M_{2}<0.05\,M_{\rm \odot}\\
\end{array}\right.
\end{equation}
where $f_{\rm dt}$ is an adjustable coefficient set to be as folllows (see Appendix\,\ref{Sect_fdt} for details).

(1) the FW+BDML model

For systems with $M_{\rm WD}^{\rm i}=0.4, 0.6,0.8\,M_{\rm \odot}$, we carry out calculations with $f_{\rm dt} =0.1$, 0.5, and 1.0; for $M_{\rm WD}^{\rm i}=1.0,1.2\,M_{\rm \odot}$, we carry out calculations with $f_{\rm dt} =0.5$ and 1.0. Then, we adopt the minimum $f_{\rm dt}$ for a specific system with which the calculation is convergent.

(2) the FW model

We adopt $f_{\rm dt}=0.5$ for all calculations.

\begin{figure*}
\centering
\includegraphics[width=15cm,height=5cm]{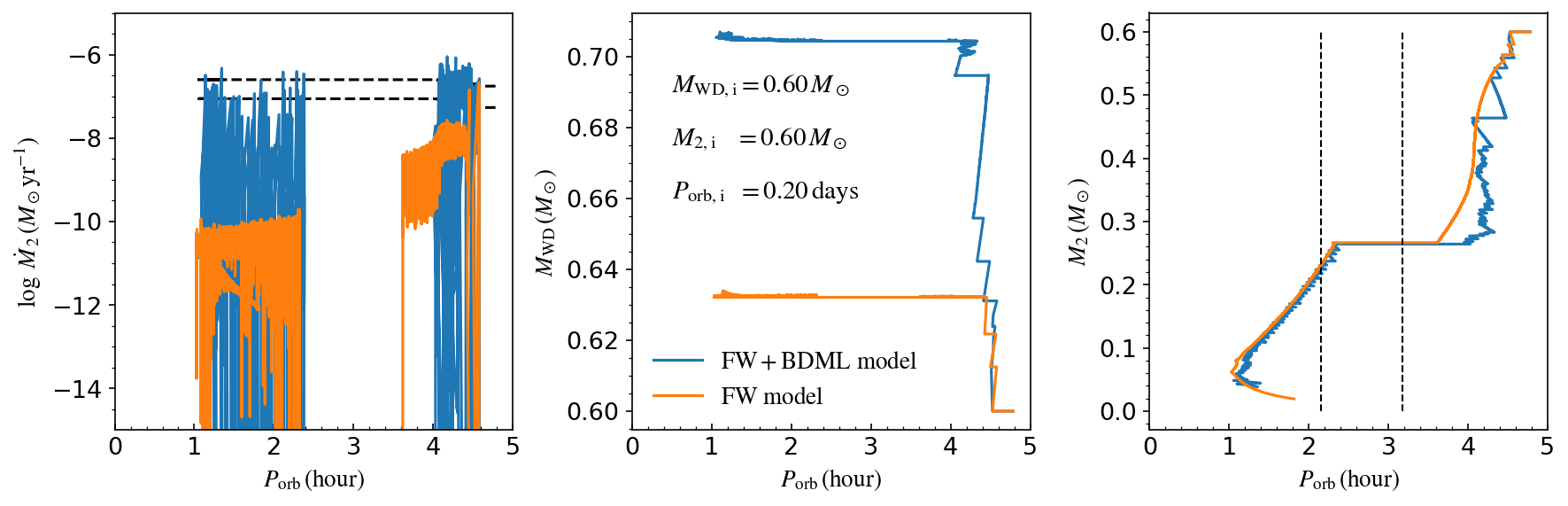}\\
\includegraphics[width=15cm,height=5cm]{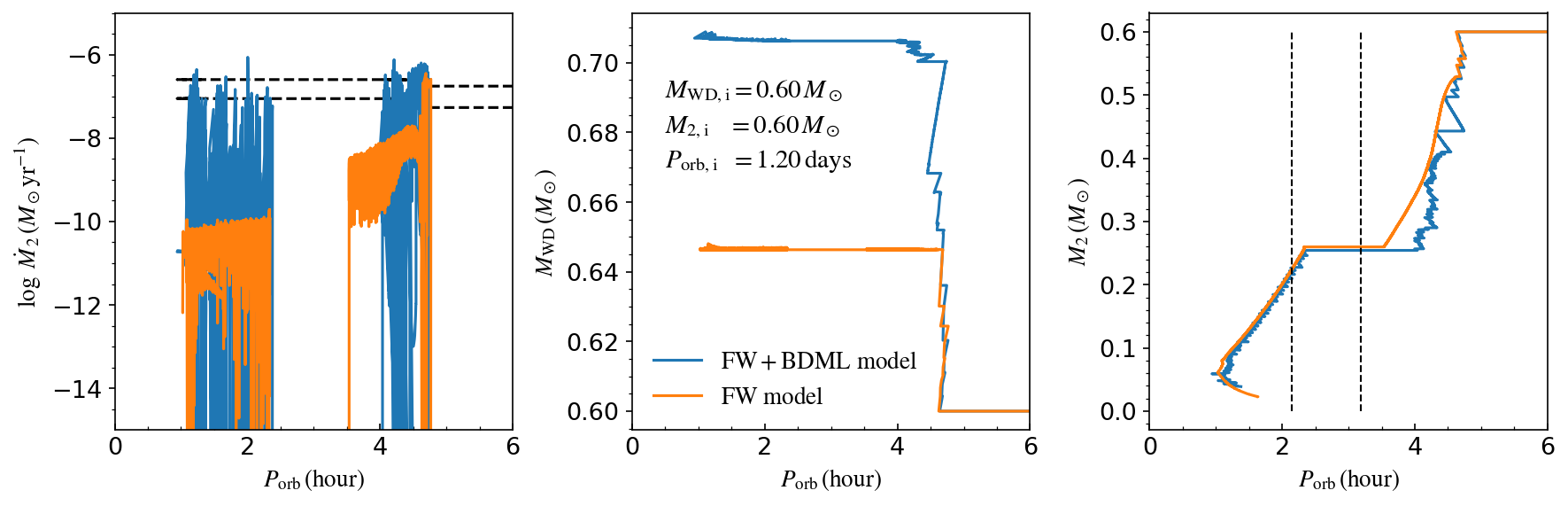}\\
\includegraphics[width=15cm,height=5cm]{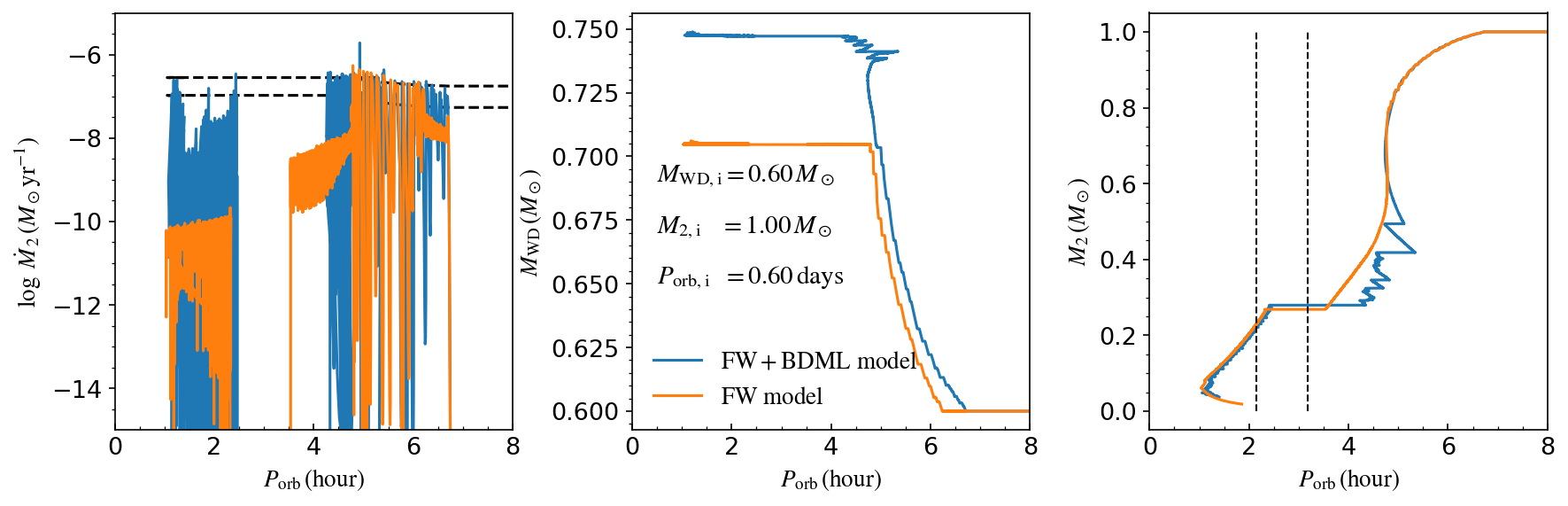}
 \caption{Evolution of CVs with a WD of initial mass $M_{\rm WD,i}=0.6\,M_{\rm \sun}$ and $f_{\rm dt}=0.5$. The left, middle, and right panels show the evolution of the mass transfer rate, the WD mass, and the companion mass as a function of the orbital period, respectively. The initial parameters are labeled in the middle panels. The blue and orange lines represent the evolutionary tracks with and without BDML considered, respectively. The dashed lines in the left panels denotes the lower and upper boundaries of the mass transfer rates for stable hydrogen burning, and in the right panels the lower (2.15\,hr) and upper boundaries (3.18\,hr) of the period gap \citep{2011ApJS..194...28K}.}
\label{Fig_MWD0.6_detailed}
\end{figure*}

 \begin{figure*}[ht]
\centering
\includegraphics[width=10cm,height=5cm]{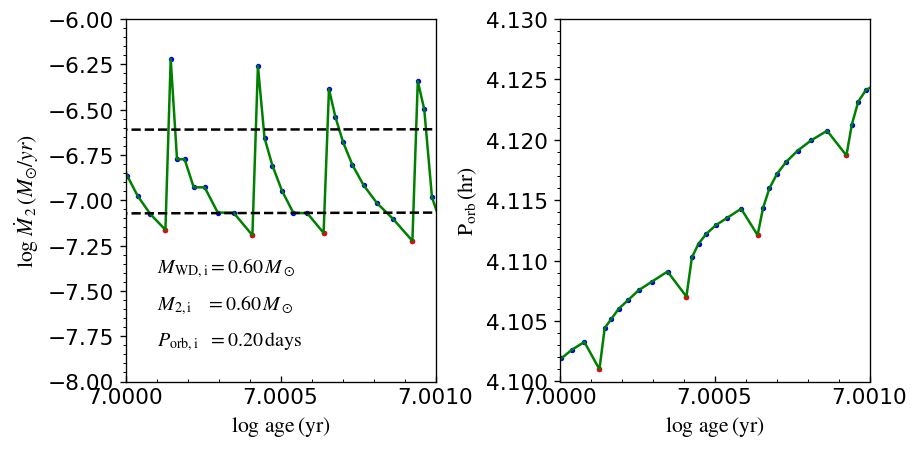}
 \caption{Detailed evolution of a CV during a few nova cycles with $M_{\rm WD,i}=0.6\,M_{\rm \odot}$. The left and right panels show the mass transfer rate and the orbital period as a function of time, respectively. In each panel, the red dots mark the location of nova eruptions, the blue dots are the evolutionary data points, linked by the green lines.}
\label{novae_parameter}
\end{figure*}

We present example evolutionary results in the models with and without BDML considered in Fig.\,\ref{Fig_MWD0.6_detailed}. The initial parameters are taken as follows: $M_{\rm WD,i}=0.6~M_{\rm \sun}$, $M_{\rm 2,i}=0.6~M_{\rm \sun}$ and $1.2~M_{\rm \sun}$, and $P_{\rm orb,i}= 0.6$ day and 1.2 days. The left, middle, and right panels show the evolution of the mass transfer rate, the WD mass, and the companion mass as a function of the orbital period, respectively. In each panel, the blue and orange lines correspond to the FW+BDML model and the FW model, respectively.  We see that in both models the evolution starts with a thermal timescale mass transfer phase, during which the mass transfer rate rises to or above the stable hydrogen burning region and the WD grows in mass to some extent (see also \citealt{2019A&A...622A..35L}).
After that the mass transfer rate decreases to  $<\dot{M}_{\rm stable}$, the binary enters the nova stage. The mass transfer proceeds discontinuously because the companion is temporarily decoupled from its Roche lobe. During this period, the mass transfer rate in the FW model remains to be $<\dot M_{\rm stable}$, while the mass transfer rate in the FW+BDML model maintains fluctuation around the stable burning region (especially when MB takes effect). 

To understand the behavior of the mass transfer rate in the BDML model, we plot the detailed CV evolution during a few nova cycles in Fig.\,\ref{novae_parameter}. 
The initial parameters are same as in the upper panel of Fig.\,\ref{Fig_MWD0.6_detailed}. 
Mass loss during the nova stages is dominantly binary-driven \citep{2022ApJ...938...31S}, leading to orbital shrink and increase in the mass transfer rate (the left and right panels, respectively). 
Because the orbital variation is taken to be instantaneous, the companion cannot immediately adjust itself to the new orbit. It slightly overfills its Roche lobe, and gives rise to increase in the mass transfer rate.
Rapid mass transfer onto the WD causes the expansion of the orbit. With decreasing angular momentum loss rate after the nova eruptions, the companion cannot keep itself to fill its Roche lobe, and the binary becomes detached. 

Back to Fig.\,\ref{Fig_MWD0.6_detailed}, the middle panel shows that the amount of mass growth in the WDs is different in the two models, because of different mass transfer trajectories.  In the FW model, the WD accumulates mass only in the initial thermal timescale mass transfer stage, the duration of which depends on the initial orbital period and mass ratio, while in the FW+BDML model, the mass growth sustains over a longer time, because the mass transfer rate can be substantially enhanced due to more angular momentum being lost during nova eruptions. For example, in the top panel the final WD masses in the FW and FW+BDML models can reach $\sim 0.63\,M_{\rm \sun}$ and $\sim 0.70\,M_{\rm \sun}$, respectively.

The orbital period evolution also shows interesting difference between the two models. They both overestimate the lower and upper edges of the period gap, compared with the observed range of $2.15-3.18$\,hr \citep{2006MNRAS.373..484K}. The FW and FW+BDML models predict the upper edge of the period gap to be $\sim 3.5$\,hr and $\sim 4$\,hr respectively, and both models predict similar lower boundary of the period gap ($\sim 2.4$\,hr). In addition, the predicted minimum orbital period is 60 min\footnote{For the default set of the atmosphere boundary conditions ``which\_atm\_option = simple\_photosphere'' in {\tt MESA}, it is difficult to evolve the systems into period bouncer in the FW+BDML model. This may be because the companion cannot adjust its radius to the new orbit due to the large variation of the orbital parameters after a nova eruption  when the companion becomes a brown dwarf. To let more systems in the FW+BDML model evolve into period bouncer, when the companion mass decreases to $0.08\,M_{\rm \odot}$, we change ``which\_atm\_option'' from the default value to ``tau\_100\_tables'' which is primarily for the evolution of low-mass stars, brown dwarfs, and planets \citep{2011ApJS..192....3P}. Note that even so, not all of the systems can evolve into the period bouncer. For this reason,  we do not provide the exact percentage of the period bouncer for population synthesis results (Sect.\,\ref{BPS-results}) because it is not accurate.}, shorter than the observed value $\sim 80$\,min \citep{2011ApJS..194...28K,2019MNRAS.486.5535M}. We will discuss the period distribution of the CV population in Fig.\,\ref{Fig-BPS_MWD_Porb_distri} below.

Finally, different from the eCAML model of \citet{2016MNRAS.455L..16S}, we do not obtain any merger outcome in the FW+BDML model.

\subsection{Comparison with observations}
\subsubsection{Angular momentum loss below the period gap}

 \begin{figure*}[ht]
\centering
\includegraphics[width=5.5cm,height=5.5cm]{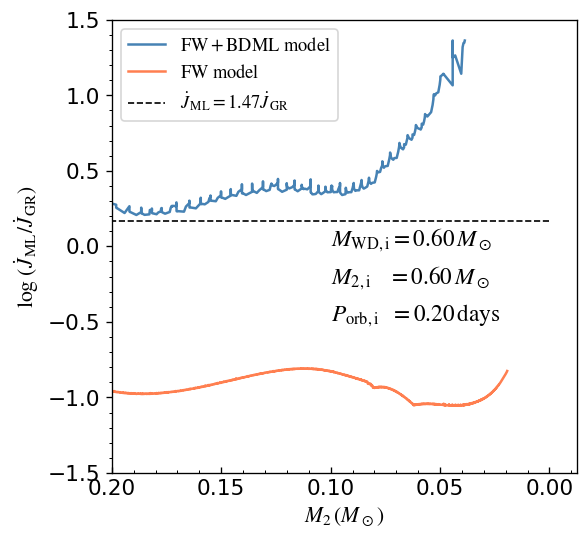}
\includegraphics[width=5.5cm,height=5.5cm]{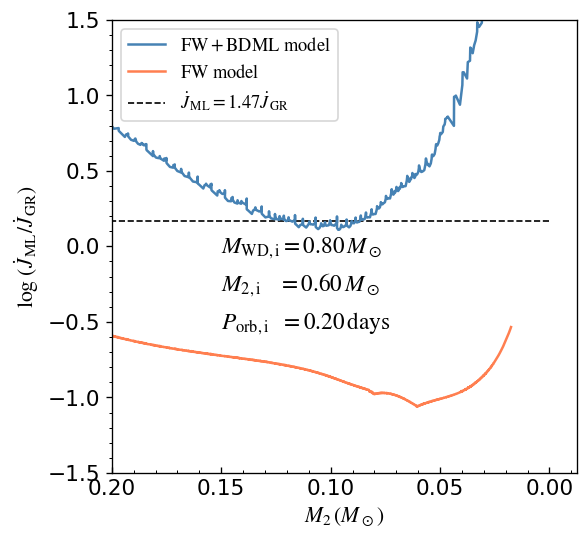}
\includegraphics[width=5.5cm,height=5.5cm]{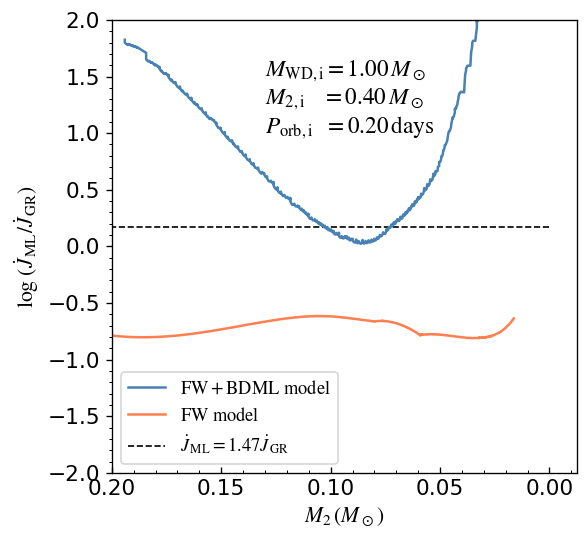}
 \caption{Ratio of $\langle \dot J_{\rm ML}\rangle$ to $\dot J_{\rm GR}$ as a function of the companion mass. The orange and blue curves stand for the FW+BDML and FW models, respectively. The dotted line denotes $\dot J_{\rm ML}=1.47\dot J_{\rm GR}$.}
\label{Fig-mean-evol}
\end{figure*}

Fig.\,\ref{Fig-mean-evol} shows the ratio of $\langle\dot J_{\rm ML}\rangle$ to $\dot J_{\rm GR}$ when $P_{\rm orb}<2$\,hr, i.e., below the period gap.
Here the time-averaged angular momentum loss rate due to mass loss is calculated by the following formula
\begin{equation}\label{Eq-jdotratio}
  \langle \dot J_{\rm ML} \rangle = \frac{\Delta J_{\rm orb}-\Delta J_{\rm GR}}{\Delta t}.
\end{equation}
then it is fitted as a function of $M_{2}$ using polynomial function with a degree of 5 via the {\tt polyfit} function in {\tt Pyhon} code.

\cite{2011ApJS..194...28K} reconstructed the complete evolutionary path followed by CVs based on the observed mass$-$radius relationship of the companion stars, and found a scaling factor for the standard angular momentum loss rates due to GR, $f_{\rm GR}=2.47\pm 0.02$ below the period gap. This means an extra angular momentum loss rate of about 1.47\,$\dot J_{\rm GR}$ being required below the period gap, which is shown with the horizontal dotted line in Fig.~\ref{Fig-mean-evol}.  We examine whether the extra angular momentum loss can be contributed by mass loss. In the FW model, $\langle \dot J_{\rm ML} \rangle/ \dot J_{\rm GR} \ll 1.47$ (shown with the yellow lines). However, $\langle \dot J_{\rm ML} \rangle/ \dot J_{\rm GR}$ is larger than 1.47 (shown with the brown lines) in the FW+BDML model. This feature suggests that BDML might serve as the origin of the extra angular momentum loss.

 \subsubsection{Binary population synthesis results}\label{BPS-results}

 \begin{figure*}[ht]
\centering
\includegraphics[width=14cm,height=8cm]{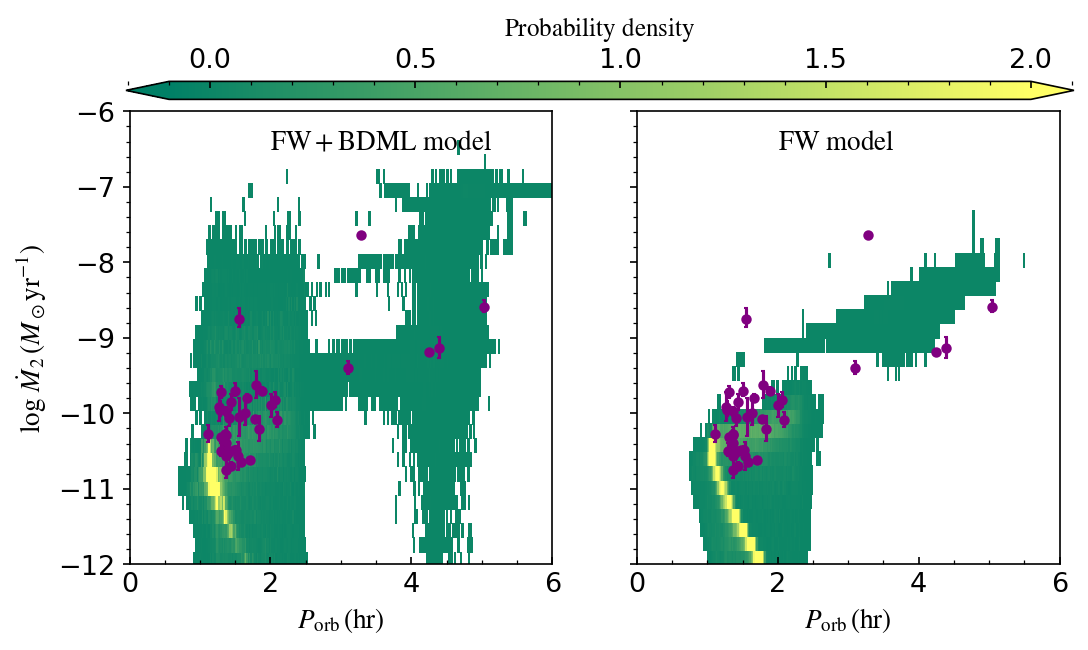}
 \caption{Comparison of the mass transfer rates in CVs between theory and observation. The left and right panels show the results in the FW+BDML model and the FW model, respectively. In each panel, the purple dots represent the observed sources with measured mean mass accretion rate from \cite{2022MNRAS.510.6110P}.}
\label{BPS_Porb_MTR}
\end{figure*}

 To investigate the overall properties of the CV population under the impact of BDML, we combine binary population synthesis with detailed evolution calculation. We firstly simulate the evolution of $10^{6}$ binaries consisting of two zero-age MS stars with Solar metallicities until the formation of WD+MS binaries by using the Monte Carlo code {\tt BSE} \citep{2000MNRAS.315..543H,2002MNRAS.329..897H}. 
Then we evolve a large number of WD+MS binaries using {\tt MESA} to get the parameter space for CVs, in which $M_{\rm WD, i}=0.4-1.2\,M_{\rm \odot}$ in steps of $0.2\,M_{\rm \odot}$, $M_{\rm 2,i}=0.1-2.0\,M_{\rm \odot}$ in steps of $0.2\,M_{\rm \odot}$, and $P_{\rm orb,i}=0.1-3.0\,\rm days$ in steps of 0.2\,day.

The key parameters for the binary population synthesis are similar as in \cite{2014ApJ...796...37S}. A constant star formation rate ($5\,M_{\rm \odot}\,\rm yr^{-1}$) is assumed over the past 10\,Gyr. The primary masses $M_{1}$ are sampled according to initial mass function of \cite{1993MNRAS.262..545K}, and the secondary masses $M_{2}$ are taken from an evenly distributed mass ratio in $[0,1]$. The initial orbits are assumed to be circular, with the binary separation in logarithm follows a flat distribution within $[3,10^{\rm 4}]\,R_{\rm \sun}$. When the primary star evolves and transfers material to the secondary via Roche lobe overflow, we assume that half of mass lost by the donor is accreted by the secondary, as suggested by \citet{2014ApJ...796...37S} due to its better reproduction of Be/X-ray binaries compared with other mass accretion efficiency recipes. If the mass transfer is dynamically unstable, common envelope evolution ensues. We take the common envelope ejection efficiency $\alpha_{\rm CE}=1/3$ \citep[e.g.,][]{2022MNRAS.513.3587Z,2023MNRAS.518.3966S}, and the results in \cite{2010ApJ...716..114X} for the binding energy parameter $\lambda$ of the envelope. We obtain the distribution of the post-common envelope WD binaries, and incorporate them with the detailed binary evolution results  calculated with {\tt MESA} to obtain the characteristics of the CV population.

It is difficult to infer the mean mass transfer rate in CVs from observations, because modern observational histories are much shorter than the evolutionary timescales. According to the compressional heating theory \citep{1995ApJ...438..876S,2003ApJ...596L.227T,2004ApJ...600..390T,2009ApJ...693.1007T}, the WDs in CVs are heated up by the compression of the accreted material across the nova cycles. Therefore, their effective temperatures ($T_{\rm eff}$) can provide an estimation of the mean mass accretion rate over the thermal timescale ($\sim 10^{3}-10^{5}$\,yr) of the WD envelope. Based on this idea, \cite{2022MNRAS.510.6110P} derived the mean mass accretion rates for 41 CVs. Comparing between theory and observation, they argued that the standard CV evolution model overestimates the mean mass accretion rate above the period gap and predicts a too narrow range of the mean mass accretion rate below the period gap.

Fig.\,\ref{BPS_Porb_MTR} displays the mass transfer rate as a function of the orbital period in the FW+BDML (left panel) and FW (right panel) models. We divide the $P_{\rm orb}-\dot{M}_2$ plane into $500\times 500$ pixels and use different colors to represent the mass transfer rates weighted by their probability densities with which  CVs appear in a specific pixel (indicated by the top color bar) \footnote{The probability density is determined by the birthrate of a source and its duration in a specific pixel in the parameter space. For technical details, see \url{https://matplotlib.org/stable/api/_as_gen/matplotlib.pyplot.hist.html}.}.   We also plot the mean mass accretion rates derived by \cite{2022MNRAS.510.6110P} with the purple dots with error bars. It is seen that the FW model covers too narrow range of the mass transfer rate both above and  below the period gap, consistent with the conclusion of \cite{2022MNRAS.510.6110P}. In comparison, the FW+BDML model can produce a wider range of the mass transfer rates, and cover almost all of the observed sources.
 
 \begin{figure*}[ht]
\centering
\includegraphics[width=16cm,height=8cm]{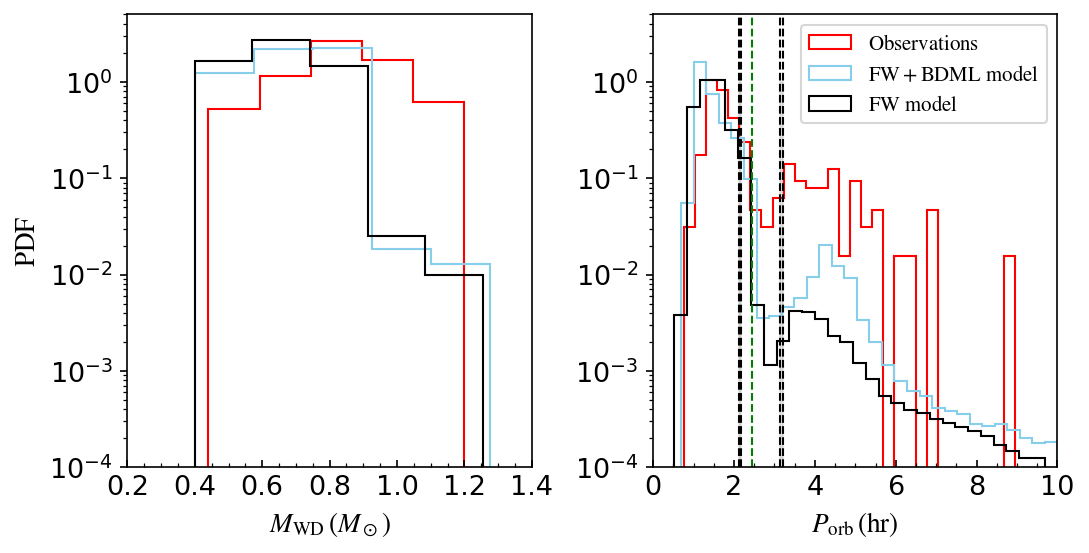}
 \caption{Distributions of the WD mass (left) and orbital period (right), compared with the observed CVs (red curves). In each panel, the black and skyblue curves correspond to the FW model and the FW+BDML model, respectively. The black dashed vertical lines in the right panel represent the lower ($2.15\pm 0.03$ hr) and upper boundaries ($3.18\pm 0.04$ hr) of the period gap given by \cite{2011ApJS..194...28K}, while the green one is the lower boundary of the gap estimated by \cite{2024A&A...682L...7S}. The data of the WD masses and orbital periods are taken from \cite{2022MNRAS.510.6110P} and \cite{2023MNRAS.524.4867I}, respectively.}
\label{Fig-BPS_MWD_Porb_distri}
\end{figure*}

 \begin{figure*}[ht]
\centering
\includegraphics[width=15cm,height=8cm]{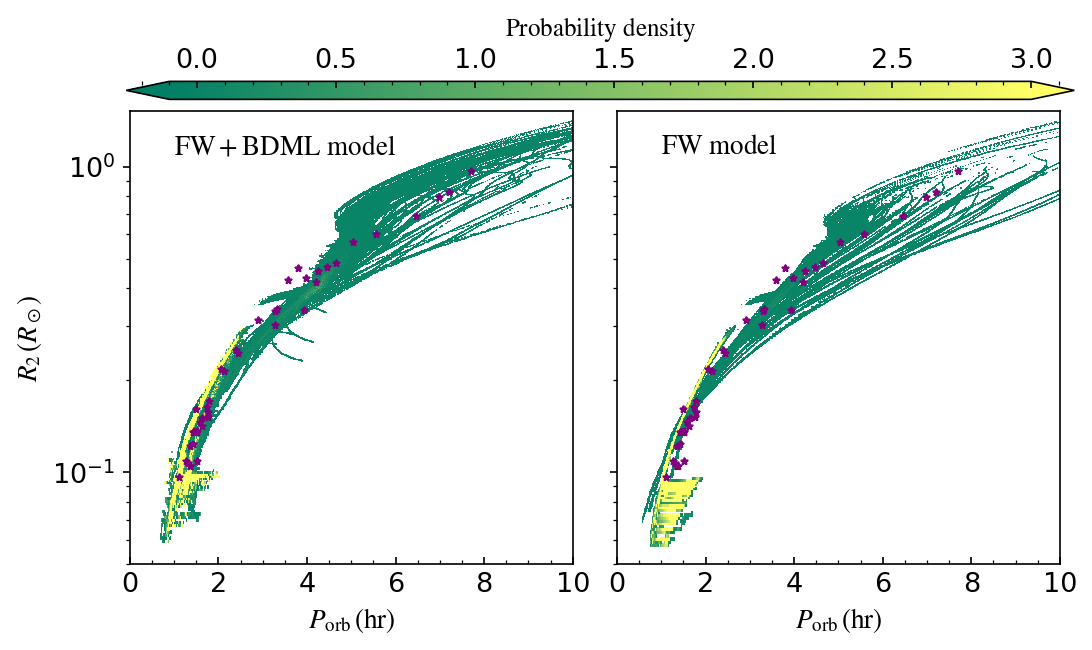}
 \caption{Distribution of CVs in the donor radius - orbital period diagram. The color bar denotes the probability density that CVs appear in a specific pixel.  The left and right panels show the results in the FW+BDML model and the FW model, respectively. The purple stars represent the observed sources from \cite{2019MNRAS.486.5535M}.}
\label{Fig-BPS_Porb_R2}
\end{figure*}

Fig.\,\ref{Fig-BPS_MWD_Porb_distri} shows the predicted WD mass (left panel) and orbital period (right panel) distributions of CVs, compared with the observations (red curves). From the left panel, we notice that although BDML shifts the WD masses to some larger values because it can drive higher mass transfer rates compared with in the FW case, there is a deficiency in massive WDs in both models compared with the observations. This demonstrates that the FW+BDML model is unable to efficiently increase the WD masses. In the right panel, both the FW and FW+BDML models  produce relatively less systems with long orbital periods compared with the observation. However, as discussed by \cite{2009MNRAS.397.2170G} and \cite{2020MNRAS.491.5717B}, the {\em Sloan Digital Sky Survey} can identify long-period CVs out to a distance $>10^4$\,pc, but closer (within $\sim 400$\,pc) for short-period CVs. This implies the observational bias toward long-period CVs and makes it difficult for proper comparison of the orbital period distributions.  The FW model seem to better reproduce the period gap, which is confined by the dotted vertical lines\footnote{It seems that \cite{2011ApJS..194...28K} slightly underestimated the lower boundary of the period gap. For example, \cite{2024A&A...682L...7S} found that the lower boundary of the period gap should be 147\,min (i.e. 2.45\,hr) which is plotted as green dashed line in Fig.\,\ref{Fig-BPS_MWD_Porb_distri}.}, while the upper boundary ($\sim 4.5$\,hr) in the FW+BDML model is longer than the observation. This stems from the fact that BDML makes the systems to sustain a high mass transfer rate, which increases the orbital period before MB stop. The width of the period gap is also dependent on the MB efficiency. \cite{2011ApJS..194...28K} found that the MB strength should be 0.66 times the most frequently cited Skumanich law \citep{1972ApJ...171..565S, 1983ApJ...275..713R}. We simulate the evolution of a systems by adopting 0.66 and 0.5 times Skumanich MB law, and find that the upper boundary of the period gap in FW+BDML model shifts from $\sim 4.5$\,hr (standard Skumanich law) to $\sim 3.7$ hr and $\sim 3.4$ hr, respectively.

Fig.\,\ref{Fig-BPS_Porb_R2} demonstrates the CV distribution in the donor radius$-$orbital period plane. We divide the parameter space into $1000 \times 1000$ pixels and use different colors to represent their number densities in a specific pixel. The most remarkable difference between the FW and FW+BDML models lies in the distribution below the period gap. CVs in the FW+BDML model can span a wider range with a relatively larger probability, and cover the majority of the observed sources, while the FW model fails.

\section{Discussion}\label{Discussion}
\subsection{Comparison with previous works}\label{Sect-diss1}
Here we compare our results with previous works which took into account the influence of nova outbursts on the evolution of CVs, including \cite{2001ApJ...563..958K}, \cite{2016ApJ...817...69N}, and \cite{2020NatAs...4..886H}.

\cite{2001ApJ...563..958K} investigated the variation of the mass transfer rate in CVs due to nova outbursts by considering the ejected material escaping directly from the surface of the WD, which is same as in our FW model. They firstly developed an analytic model for the mass transfer rate spectrum of CVs and obtained the nova-induced width of the mass transfer rate spectrum. Then they performed population synthesis simulation to explore how broad the range in the mass transfer rate would be for an observed collection of CVs with different binary parameters. They concluded that nova outbursts can give rise to the variation of the mass transfer rate by a factor 10 above the period gap, especially for period $\sim 3-6\,\rm hr$, but this is reached only when the nova-triggering mass is larger than a few $10^{-4} M_{\rm \odot}$. In addition, nova-induced broadening of the mass transfer rate is insignificant for CVs below the period gap or with high-mass WDs.

Our calculations for the FW model show  similar results with \cite{2001ApJ...563..958K}. The right panel of Fig.\,\ref{BPS_Porb_MTR} demonstrates that the mass transfer rate spreads over $1-2$ orders of magnitude above the period gap, while the width of the mass transfer rates above and below the period gap is still relatively narrow in comparison with observations. However, for the FW+BDML model, the mass transfer rates are distributed over a larger range both above and below the period gap, and can better cover the range of the observed mass transfer rate of CVs.

\cite{2016ApJ...817...69N} investigated the influence of nova outbursts on the formation and evolution of CVs by assuming a portion of the nova ejecta forms a common envelope-like structure around the binary and interacts with the companion. In their work, the angular momentum is finally extracted via a “circumbinary” ring. Different from our calculations considering instantaneous mass ejection, they modeled the mass loss process as a continuous process, and calculated the stability of the mass transfer with common envelope-like mass ejection. They found that CVs are more likely to merge with the increasing fraction of mass ejection. In order to let this mechanism work comfortably, there should be $\sim 40\%$ nova ejecta being lost via common envelope in CVs with low-mass WDs, and it is lower for more massive WDs. This tendency is qualitatively consistent with the assumption adopted in our work \citep{2022ApJ...938...31S}.

In the BDML model, a large fraction of the ejected mass leaves the binary through the $L_2$ point, but we do not find any merger outcome. The specific orbital angular momentum of the mass lost through circumbinary ring is $j_{\rm CB}=-1.5a^{2}\omega$ \citep{1997A&A...327..620S,2012ApJ...745..165S,2016ApJ...817...69N}, which is comparable with that in the case of overflow through the $L_2$ point, $j_{\rm L_{2}}=-a_{\rm L_{2}}^{2}\omega$ for CVs. 
The main reason for the difference may stem from the different treatments of angular momentum loss below and above the stable burning region. \cite{2016ApJ...817...69N} assumed the same angular momentum loss form below and above $\dot M_{\rm stable}$, which ensures a large angular momentum loss rate even when $|\dot M_{2}|>\dot M_{\rm stable}$, while in our work, if nova eruptions causes $\dot M_{2}$ to increase beyond $\dot M_{\rm crit}$, the excess material is assumed to be lost as traditional fast wind. This switch results in a drop in the angular momentum loss rate, which in turn leads to the decrease in $\dot M_{2}$. Therefore, nova eruptions do not lead to runaway mass transfer in our case.

\cite{2020NatAs...4..886H} simulated the multi-Gyr evolution of nova systems by modelling every eruption’s thermonuclear runaway, mass and angular momentum losses, feedback due to irradiation and variable mass transfer rate, and orbital period changes. They showed that 
feedback-dominated evolution can reproduce the observed range of the mass transfer rates at a given orbital period, with large and cyclic kyr–Myr timescale changes. In short-period binaries the mass transfer rate evolution also demonstrates Myr-long deep hibernation.
This interprets why most known novae are observed to have orbital periods longer than 3\,hr.

Similar to \cite{2020NatAs...4..886H}, we also observe cyclic variation of the accretion rate onto the WDs, but with different origins. Our FW+BDML model predicts a wide range of $\dot M_{2}$ at all orbital periods. In \cite{2020NatAs...4..886H}, the enhanced Roche lobe overflow is caused by bloating of the companion star which is strongly irradiated by the WD, while in our work, this is caused by orbital shrink due to mass and angular momentum loss during nova eruptions. Our calculations indicate that CVs can enter deep hibernation. However, different from  \cite{2020NatAs...4..886H} in which deep hibernation only exists in short-period systems as the effect of the cooling WD’s irradiation diminishes, both short- and long-period systems can hibernate in our calculations, especially in the FW+BDML model, because of the widening of the orbital period.

\subsection{Orbital period variations due to nova eruptions}

\begin{figure*}[ht]
\centering
\includegraphics[width=9cm,height=9cm]{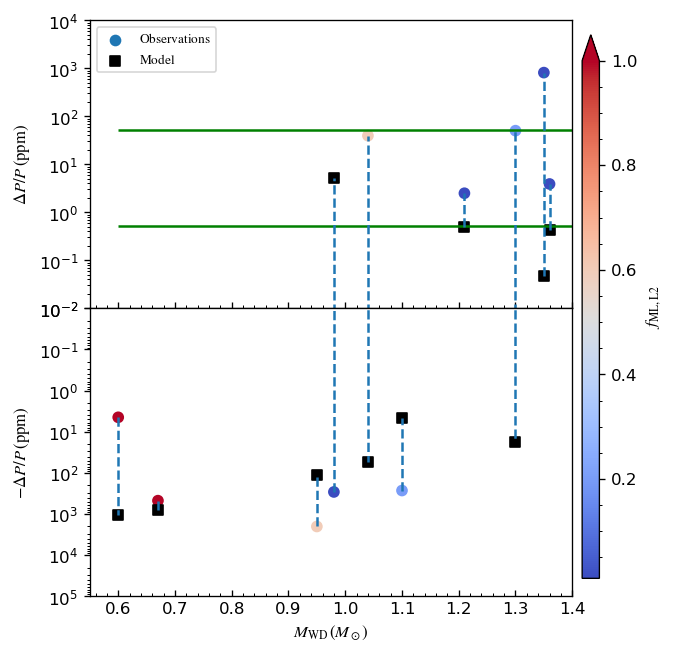}
\caption{Comparison of the measured $\Delta P/P$  (colored circles) in the unit of parts-per-million (ppm) and theoretical predictions (black squares). The color bar represents the inferred mass loss fraction in the form of BDML. In the upper and low panels the $\Delta P$ values are positive and negative, respectively. The two green lines in the upper panel represent the predictions in the FW model with $\Delta M_{\rm ign}=10^{-6}$ and $10^{-4}\,M_{\rm \odot}$, respectively. The dashed lines connect the observed and theoretical values for given sources. The observed data are taken from \cite{2023MNRAS.525..785S}.}
\label{Fig-Dp-model-and-Obser}
\end{figure*}

The variations $\Delta P$ of the orbital period (or binary separation) across nova eruptions can be taken as a probe of the angular momentum loss mechanism associated with mass loss. The traditional assumption of the fast wind always predicts a positive $\Delta P$ (see Eq.\,[\ref{Eq-dota-FW}]). \cite{2023MNRAS.525..785S} measured the orbital period changes for 14 novae, and found that near half of the novae have negative $\Delta P$, which is inconsistent with the fast wind assumption. He proposed that the most likely mechanism that leads to the negative $\Delta P$ is the asymmetric mass ejection which exerts a kick on the WD (See also \citealt{2020MNRAS.492.3343S}).

Here we examine the orbital period change caused by BDML. In Fig.\,\ref{Fig-Dp-model-and-Obser} the circles represent measured $\Delta P/P$  by \cite{2023MNRAS.525..785S} as a function of the WD mass, and the black squares represent the predicted  $\Delta P/P$ at the same WD mass and orbital period in the FW+BDML model. The color represents the inferred fraction of mass loss through the $L_2$ point outflow (i.e. $f_{\rm ML, L2}$)\footnote{We adopt the observed $M_{\rm WD}$ and masses of nova ejection from Table 10 of \cite{2023MNRAS.525..785S}. Combining with Fig.\,8 of \cite{2022ApJ...938...31S}, we can obtain  $f_{\rm ML, L2}$.}. The FW model always predicts positive $\Delta P$, as depicted by the two green lines with $\Delta M_{\rm ign}=10^{-6}$ and $10^{-4}\,M_{\rm \odot}$.
Fig.\,\ref{Fig-Dp-model-and-Obser} shows that WDs with masses above and below $\sim 1.0-1.1\,M_{\rm \odot}$ seem to have positive and negative $\Delta P$, respectively. However, it is not clear whether this tendency remains if more data is collected in the future. The colors of the circles indicate that the fraction of BDML decreases with increasing WD mass, in agreement with \cite{2022ApJ...938...31S}. 
Our predicted $\Delta P$s can roughly cover the observational range. However, for systems with massive WDs, our calculations underestimate $\Delta P$. In a few cases, the predicted $\Delta P$s have opposite signs compared with the observational data, suggesting that the mass loss alone is insufficient to account for the observed $\Delta P$. Probably  other mechanisms like nova-induced kicks  \citep{2023MNRAS.525..785S} might be required.

\subsection{Peculiar supersoft X-ray sources}\label{Explain-SSSs}

Our calculations show that BDML can induce rapid mass transfer, so the binaries may switch between CVs and supersoft X-ray sources (SSSs). SSSs are X-ray sources with blackbody temperature of 20$-$100 eV and luminosity of $\sim 10^{35}-10^{38}$\,erg\,s$^{-1}$ \citep{KH97}. The supersoft X-ray is suggested to come from stable nuclear burning on the surface of WDs, which requires a mass transfer rate of $|\dot M_{2}|\gtrsim 10^{-7}\,M_{\rm \odot}\,\rm yr^{-1}$. In the standard picture the WD is accreting from a more massive main sequence companion star, so the mass transfer proceeds on a (sub)thermal timescale \citep{1992A&A...262...97V}. Because mass is transferred from the more massive donor star to the less massive WD, the orbit period is expected to decrease with time.

\begin{figure*}[ht]
\centering
\includegraphics[width=16cm,height=5.5cm]{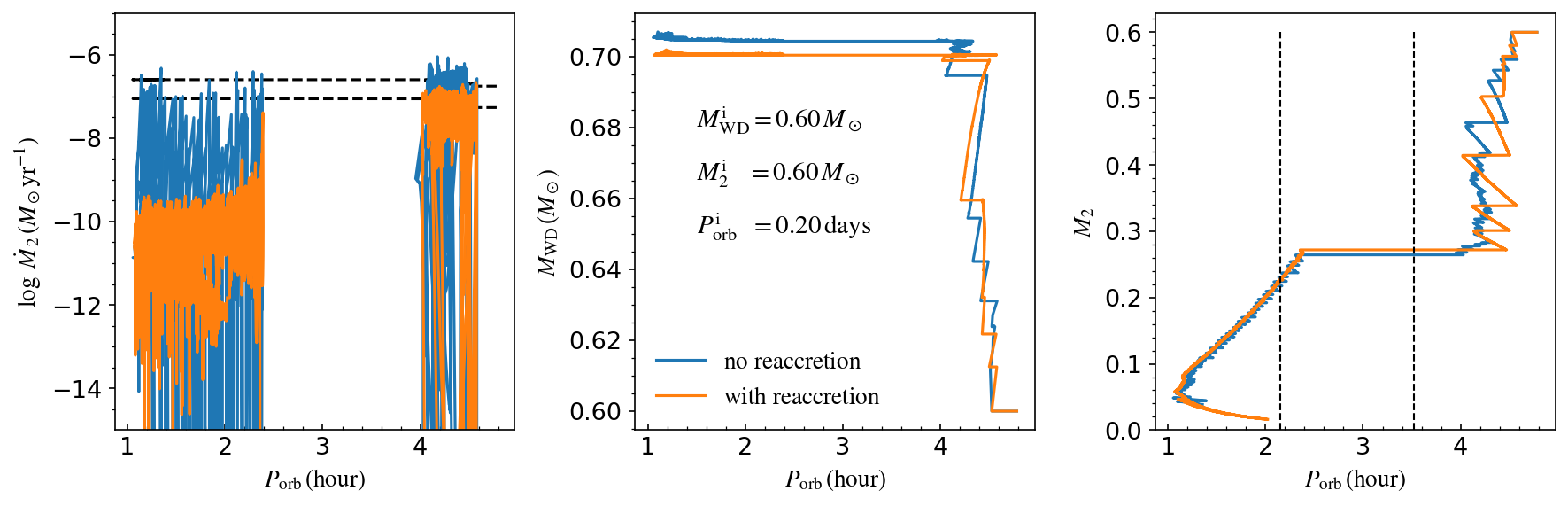}
 \caption{Example evolution that displays the influence of re-accretion of the nova ejecta by the companion with $\gamma_{\rm re}=0.5$. The steelblue and orange lines show the cases with $\gamma_{\rm re}=0$ and  0.5, respectively.}
\label{Fig-with_reacc}
\end{figure*}

CAL 87 is thought to be the prototype of SSSs \citep{1990ApJ...350..288C}. However, the WD mass ($\sim 1.35\,M_{\rm \odot}$) is significantly larger than the companion mass ($\sim 0.34\,M_{\rm \odot}$) \citep{2004ApJ...612L..53S,2007A&A...472L..21O}. Moreover, it has an orbital period of $10.6$\,hr, which is increasing at a rate of $\sim 6\times 10^{-10}\rm s\,s^{-1}$ \citep{1997MNRAS.287..699A, 2014ApJ...792...20R}.  
1E 0035.4$-$7230 and RX J0537.7$-$7034 are other two peculiar SSSs, which share the  similar properties: (i) the companion masses ($\sim 0.4\,M_{\rm \odot}$) are lower than the WD masses \citep[$\sim 0.6-0.7\,M_{\rm \odot}$,][]{1999A&A...346..453K,2000A&A...355.1041G}; (ii) the orbital orbital periods are very short \citep[$\sim 4$\,hr,][]{1997ApJ...489..903C,2000A&A...355.1041G} which also indicate very low-mass donor stars. The small donor mass and orbital period increase are in contradiction with the expectation of the standard model.
An excited-wind-driven mass-transfer model was suggested by \cite{2015ApJ...815...17A} to explain these peculiarities. In this model, irradiation of the companion star by the soft X-rays from the WD excites strong winds that drive the binary to sustain a high mass transfer rate \citep{1998A&A...338..957V}. Here we propose that BDML may serve as an alternative mechanism to explain the formation of such SSSs.
Figs.\,\ref{Fig_MWD0.6_detailed} and \ref{BPS_Porb_MTR} show that some CVs experience rapid mass transfer episodes, during which the observed mass transfer rates, companion masses, orbital periods and their changing rates can be accounted for.

\subsection{The effect of re-accretion}\label{Sect-re-acc}

In our calculations we assume no re-accretion of the nova ejecta by the companion star, i.e., $\gamma_{\rm re}=0$. However, in the case of slow wind re-accretion should be possible. To examine its effect we change $\gamma_{\rm re}$ to be 0.1 and 0.5. In the former case we do not find any remarkable changes in the results. Fig.\,\ref{Fig-with_reacc} shows an example of evolution with $\gamma_{\rm re}=0.5$. The most apparent feature is that the mass transfer rate evolution becomes smoother with re-accretion considered, especially after MB stops. As mentioned earlier, due to rapid BDML-induced orbit expansion, the companion star is unable to adjust itself with the new orbit, and become detached from its the Roche lobe. However, re-accretion makes this tendency weaker. Therefore, the binary shows a relatively small variation in the mass transfer rate when re-accretion is considered.

Re-accretion has minor influence on the WD mass growth, period gap, and minimum orbital period. However, a systematic research should be made to examine the effect of re-accretion on the population properties of CVs, which is beyond the scope of this paper.

\section{Summary}\label{Sect-Con-Diss}

Our understanding of the CV evolution is far from being complete. The standard theories have difficulties in explaining the observed properties of the CV population, including the mean WD mass, the space density, and the absence of period bouncer systems \citep{2016MNRAS.455L..16S,2018MNRAS.478.5626B,2020MNRAS.494.3799P}. One possible solution is that MB continues to operate even when the companion star becomes fully convective.
An alternative mechanism is eCAML, with which it is claimed that almost all of the discrepancies between the theory and observation can be resolved \citep{2016MNRAS.455L..16S,2017MNRAS.466L..63Z}.
However, the physical mechanism behind eCAML is not yet clear.
In this paper, we explore the impact of BDML on the evolution of CVs based on recent hydrodynamical simulations of classical nova outflows \citep{2022ApJ...938...31S}, which showed that a large fraction ($f_{\rm ML, L2}$)  of the accreting material is lost through the outer Lagrange point $L_{2}$. 
We consider instantaneous mass ejection during each nova eruption rather than continuous mass loss usually adopted in previous works \citep[e.g. ][]{2016ApJ...833...83K}. We combine both detail binary evolution calculation with {\tt MESA} and binary population synthesis calculation with {\tt BSE} to follow the formation and evolution of CVs.  Comparing the results with the observed CV population, we reach the following conclusions.

(1) Considering BDML can significantly influence the mass transfer processes in CVs. The mass transfer proceeds discontinuously with episodes of enhanced mass transfer rates. The mass transfer rates at given orbital periods occupy a large range compared with predicted by the traditional FW model. This provides a possible explanation for the observed mass transfer rate distribution in CVs.

(2) We do not find runaway mass transfer process and a merger outcome for binaries with low mass WDs ($M_{\rm WD,i}\le 0.6\,M_{\rm \odot}$) in the  FW+BDML model. This is because different angular momentum loss mechanisms take action when the binary switches between the nova stage and the stable burning stage. This limits the mass transfer rates not to exceed the critical accretion rate for stable burning by a large factor. Consequently, the WDs can grow mass by at most $\sim 0.1\,M_{\sun}$ during the whole CV evolution.

(3) Our results reveal that the rate of angular momentum loss caused by BDML is significantly higher than that by FW. Its magnitude depends on the initial parameters of the CVs and their evolutionary stage, and it seems that BDML could provide extra angular momentum loss below the period gap found by \cite{2011ApJS..194...28K}.




(4) the FW+BDML model predicts both positive and negative variations of orbital period $\Delta P$ across nova eruptions, which are roughly in line with recent observations by \cite{2023MNRAS.525..785S}.

(5) Our results also suggest that the FW+BDML model may potentially interpret the properties of some peculiar supersoft X-ray sources, e.g. CAL 87, 1E 0035.4$-$7230 and RX J0537.7$-$7034.

We conclude that more systematic investigations taking into account BDML as well as the feedback of irradiation of the companion star and re-accretion of the companion star are required, to understand the formation and evolution of CVs.

\section*{Acknowledgments}
\quad We are grateful to the referee for insightful comments that helped improve the manuscript. This work was supported by the National Key Research and Development Program of China (2021YFA0718500), the Natural Science Foundation of China under grant No. 12041301 and 12121003.

\section*{Data Availability}
\quad All data underlying this article will be shared on reasonable request to the corresponding authors.

\begin{appendix}
\renewcommand\thefigure{\Alph{section}\arabic{figure}} 
\renewcommand\thetable{\Alph{section}\arabic{table}}

\section{The percentage of mass loss from the $L_2$ point}\label{Appenx-f-L2}
We calculate $f_{\rm ML,L2}$ for different WDs from the following relation,
\begin{equation}\label{Eq-f-L2}
f_{\rm ML,L2} =\left\{
 \begin{array}{ll}
  1, ~~{\rm for}~M_{\rm WD}<0.7\,M_{\rm \odot}\\
f_{\rm ML,L2}\,(M_{\rm WD}=0.8\,M_{\rm \odot}), ~~{\rm for}~M_{\rm WD}\in [0.7,0.9)\,M_{\rm \odot}\\
f_{\rm ML,L2}\,(M_{\rm WD}=1.0\,M_{\rm \odot}), ~~{\rm for}~M_{\rm WD}\in [0.9,1.1)\,M_{\rm \odot}\\
f_{\rm ML,L2}\,(M_{\rm WD}=1.2\,M_{\rm \odot}), ~~{\rm for}~M_{\rm WD}\in [1.1,1.3]\,M_{\rm \odot}.\\
\end{array}\right.
\end{equation}
Here $f_{\rm ML,L2}$ is extracted from Fig. 8 in \cite{2022ApJ...938...31S}. The data are listed in Table\,\ref{Tab-fL2}. 

\setcounter{table}{0}
\begin{table*}\label{Tab-fL2}
 \begin{minipage}[t]{1\textwidth}
  \centering
 \caption{The values of $f_{\rm ML,L2}$ for different $M_{\rm WD}$ and $\Delta M_{\rm ign}$.}
   \centering
\begin{tabular}{|p{0.16\textwidth}|p{0.16\textwidth}|p{0.16\textwidth}|p{0.16\textwidth}|p{0.16\textwidth}|p{0.16\textwidth}|}
    \hline
    \multicolumn{2}{|c|}{$M_{\rm WD}=0.8\,M_{\rm \odot}$} & \multicolumn{2}{|c|} {$M_{\rm WD}=1.0\,M_{\rm \odot}$} & \multicolumn{2}{|c|}{$M_{\rm WD}=1.2\,M_{\rm \odot}$} \\
    \hline
    $\Delta M_{\rm ign}\,(M_{\rm \odot})$ & $f_{\rm ML,L2}$ & $\Delta M_{\rm ign}\,(M_{\rm \odot})$ & $f_{\rm ML,L2}$ & $\Delta M_{\rm ign}\,(M_{\rm \odot})$ & $f_{\rm ML,L2}$\\
     \hline
   3.00000e-05 & 0.00000 & 1.00000e-05 & 0.00000 & 3.00000e-06 & 0.00000 \\
   \hline
   3.54005e-05 & 0.06866 & 1.18002e-05 & 0.06031 & 3.54010e-06 & 0.03472 \\
   \hline
   4.08011e-05 & 0.13733 & 1.36004e-05 & 0.12061 & 4.08010e-06 & 0.06944 \\
   \hline
   4.62016e-05 & 0.20599 & 1.54005e-05 & 0.18092 & 4.62020e-06 & 0.10415 \\
   \hline
   5.16022e-05 & 0.27466 & 1.72007e-05 & 0.24122 & 5.16020e-06 & 0.13887 \\
   \hline
   5.70027e-05 & 0.34332 & 1.90009e-05 & 0.30153 & 5.70030e-06 & 0.17359 \\
   \hline
   6.24032e-05 & 0.41198 & 2.08011e-05 & 0.36184 & 6.24030e-06 & 0.20831 \\
   \hline
   6.78038e-05 & 0.48065 & 2.26013e-05 & 0.42214 & 6.78040e-06 & 0.24302 \\
   \hline
   7.32043e-05 & 0.54931 & 2.44014e-05 & 0.48245 & 7.32040e-06 & 0.27774 \\
   \hline
   7.86049e-05 & 0.61798 & 2.62016e-05 & 0.54275 & 7.86050e-06 & 0.31246 \\
   \hline
   8.40054e-05 & 0.68664 & 2.80018e-05 & 0.60306 & 8.40050e-06 & 0.34718 \\
   \hline
   8.94059e-05 & 0.75530 & 2.98020e-05 & 0.66337 & 8.94060e-06 & 0.38190 \\
   \hline
   9.48065e-05 & 0.82397 & 3.16022e-05 & 0.67595 & 9.48060e-06 & 0.41661 \\
   \hline
   1.00207e-04 & 0.89008 & 3.34023e-05 & 0.68264 & 1.00207e-05 & 0.45041 \\
   \hline
   1.05608e-04 & 0.89224 & 3.52025e-05 & 0.68932 & 1.05608e-05 & 0.46122 \\
   \hline
   1.11008e-04 & 0.89440 & 3.70027e-05 & 0.69601 & 1.11008e-05 & 0.47202 \\
   \hline
   1.16409e-04 & 0.89656 & 3.88029e-05 & 0.70270 & 1.16409e-05 & 0.48282 \\
   \hline
   1.21809e-04 & 0.89872 & 4.06031e-05 & 0.70938 & 1.21809e-05 & 0.49362 \\
   \hline
   1.27210e-04 & 0.90088 & 4.24032e-05 & 0.71607 & 1.27210e-05 & 0.50442 \\
   \hline
   1.32610e-04 & 0.90304 & 4.42034e-05 & 0.72276 & 1.32610e-05 & 0.51522 \\
   \hline
   1.38011e-04 & 0.90520 & 4.60036e-05 & 0.72944 & 1.38011e-05 & 0.52602 \\
   \hline
   1.43411e-04 & 0.90736 & 4.78038e-05 & 0.73613 & 1.43411e-05 & 0.53682 \\
   \hline
   1.48812e-04 & 0.90952 & 4.96040e-05 & 0.74281 & 1.48812e-05 & 0.54762 \\
   \hline
   1.54212e-04 & 0.91168 & 5.14041e-05 & 0.74950 & 1.54212e-05 & 0.55842 \\
   \hline
   1.59613e-04 & 0.91385 & 5.32043e-05 & 0.75619 & 1.59613e-05 & 0.56923 \\
   \hline
   1.65013e-04 & 0.91601 & 5.50045e-05 & 0.76287 & 1.65014e-05 & 0.58003 \\
   \hline
   1.70414e-04 & 0.91817 & 5.68047e-05 & 0.76956 & 1.70414e-05 & 0.59083 \\
   \hline
   1.75815e-04 & 0.92033 & 5.86049e-05 & 0.77625 & 1.75815e-05 & 0.60163 \\
   \hline
   1.81215e-04 & 0.92249 & 6.04050e-05 & 0.78293 & 1.81215e-05 & 0.61243 \\
   \hline
   1.86616e-04 & 0.92465 & 6.22052e-05 & 0.78962 & 1.86616e-05 & 0.62323 \\
   \hline
   1.92016e-04 & 0.92681 & 6.40054e-05 & 0.79631 & 1.92016e-05 & 0.63403 \\
   \hline
   1.97417e-04 & 0.92897 & 6.58056e-05 & 0.80299 & 1.97417e-05 & 0.64483 \\
   \hline
   2.02817e-04 & 0.93113 & 6.76058e-05 & 0.80968 & 2.02817e-05 & 0.65563 \\
   \hline
   2.08218e-04 & 0.93329 & 6.94059e-05 & 0.81636 & 2.08218e-05 & 0.66644 \\
   \hline
   2.13618e-04 & 0.93545 & 7.12061e-05 & 0.82305 & 2.13618e-05 & 0.67724 \\
   \hline
   2.19019e-04 & 0.93761 & 7.30063e-05 & 0.82974 & 2.19019e-05 & 0.68804 \\
   \hline
   2.24419e-04 & 0.93977 & 7.48065e-05 & 0.83642 & 2.24419e-05 & 0.69884 \\
   \hline
   2.29820e-04 & 0.94193 & 7.66067e-05 & 0.84311 & 2.29820e-05 & 0.70964 \\
   \hline
   2.35220e-04 & 0.94409 & 7.84068e-05 & 0.84980 & 2.35221e-05 & 0.72044 \\
   \hline
   2.40621e-04 & 0.94625 & 8.02070e-05 & 0.85648 & 2.40621e-05 & 0.73124 \\
   \hline
   2.46022e-04 & 0.94841 & 8.20072e-05 & 0.86317 & 2.46022e-05 & 0.74204 \\
   \hline
   2.51422e-04 & 0.95057 & 8.38074e-05 & 0.86986 & 2.51422e-05 & 0.75284 \\
   \hline
   2.56823e-04 & 0.95273 & 8.56076e-05 & 0.87654 & 2.56823e-05 & 0.76365 \\
   \hline
   2.62223e-04 & 0.95489 & 8.74077e-05 & 0.88323 & 2.62223e-05 & 0.77445 \\
   \hline
   2.67624e-04 & 0.95705 & 8.92079e-05 & 0.88992 & 2.67624e-05 & 0.78525 \\
   \hline
   2.73024e-04 & 0.95921 & 9.10081e-05 & 0.89660 & 2.73024e-05 & 0.79605 \\
   \hline
   2.78425e-04 & 0.96137 & 9.28083e-05 & 0.90329 & 2.78425e-05 & 0.80685 \\
   \hline
   2.83825e-04 & 0.96353 & 9.46085e-05 & 0.90997 & 2.83825e-05 & 0.81765 \\
   \hline
   2.89226e-04 & 0.96569 & 9.64086e-05 & 0.91666 & 2.89226e-05 & 0.82845 \\
   \hline
   2.94626e-04 & 0.96785 & 9.82088e-05 & 0.92335 & 2.94626e-05 & 0.83925 \\
   \hline
\end{tabular}
\end{minipage}
\end{table*}

\section{Influence of the timestep}\label{Sect_fdt}%

To examine the influence of nova eruptions on the evolution of CVs, the timestep in the calculation should be short enough ($< \tau_{\rm rec}$) to resolve the nova eruptions. In {\tt MESA}, the timestep is limited to {\it max\_timestep}, which is adjusted automatically if  {\it max\_timestep} $\le 0$. We let the timestep control to be
\begin{equation}\label{Eq-basic-dt}
max\_timestep =\left\{
 \begin{array}{ll}
  f_{\rm dt}\tau_{\rm rec}, & {\rm if}~|\dot M_{2}|<\dot M_{\rm stable}\\
 0, & {\rm if}~|\dot M_{2}|\ge \dot M_{\rm stable}~{\rm or}~M_{2}<0.05\,M_{\rm \odot}\\
\end{array}\right.
\end{equation}
where $f_{\rm dt}$ is an adjustable coefficient in [0,1]. We let $f_{\rm dt}=0$ for the region of $ |\dot M_{2}|\ge \dot M_{\rm stable}$  or $M_{2}<0.05\,M_{\rm \odot}$ because the systems in the former region do not experience nova eruptions and systems  in the latter region will evolve to period bouncer thus their evolutionary timescale is long.

Fig.\,\ref{Fig-compare-fdt} presents an example of the influence of timestep control on the evolution of CVs in the FW+BDML model with $M_{\rm WD,i}=0.6\,M_{\rm \odot}$, where $f_{\rm dt}$ takes the values of 0.03, 0.05, 0.1 and 1.0. Smaller $f_{\rm dt}$ like 0.01 would make the calculation be non-convergent for almost all of systems because of too small timestep. In the $\dot M_{2}-M_{2}$ plane, we see that $\dot M_{2}$ can reach larger value for smaller $f_{\rm dt}$. In the $P_{\rm orb}-M_{2}$ plane, the evolutions are roughly similar when $f_{\rm dt}<0.1$. In addition, the  $P_{\rm orb}(M_{2})$ evolution shows that MB stops at smaller $M_{2}$ with $f_{\rm dt}=0.03$ than with larger $f_{\rm dt}$. We also see from the $M_{\rm WD}-M_{2}$ plane that changes in $f_{\rm dt}$ slightly influence the growth of WD mass which saturates when $f_{\rm dt}\le 0.05$. Therefore, in principle, the results should be more precise if smaller $f_{\rm dt}$ is adopted. However, a small $f_{\rm dt}$ may cause the non-convergence of calculation, for example, the calculations become non-convergent when $M_{2}<0.1\,M_{\rm \odot}$ for $f_{\rm dt}=0.03$ and 0.1.

\setcounter{figure}{0}
\begin{figure*}[ht]
\centering
\plotone{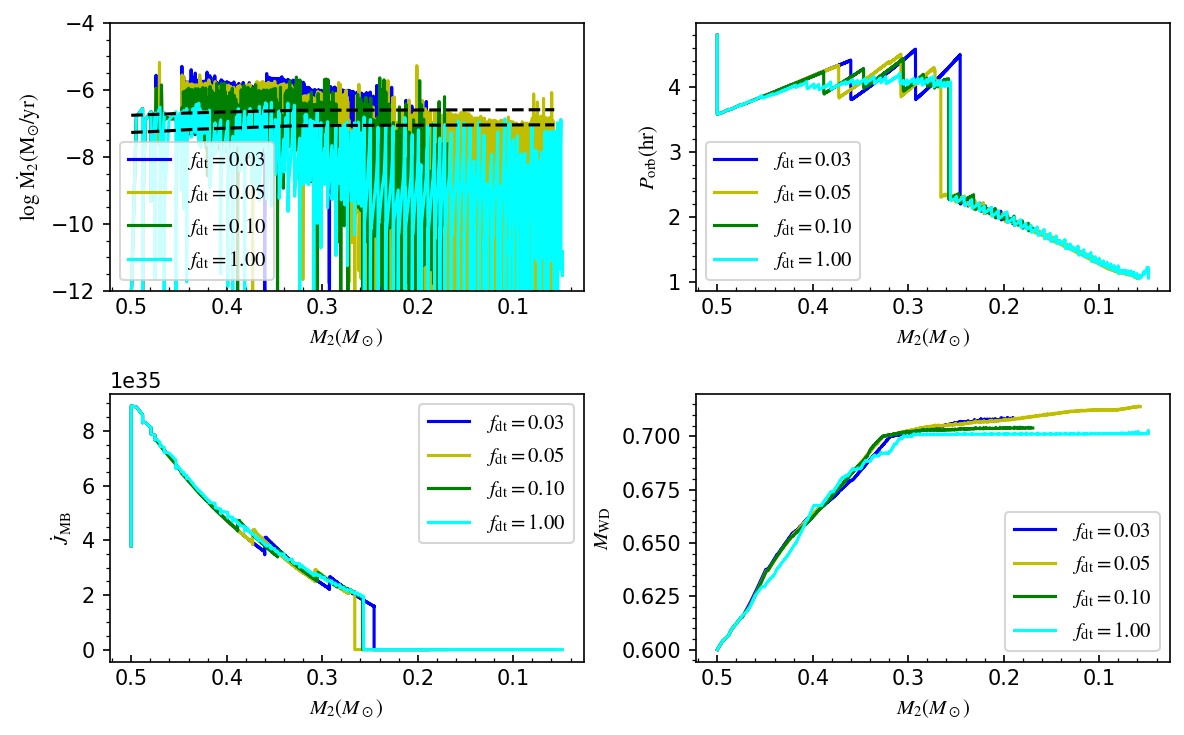}
 \caption{Influence of $f_{\rm dt}$ on the evolution of CVs in the FW+BDML model. The initial parameters are $M_{\rm WD,i}=0.6\,M_{\rm \odot}$, $M_{\rm 2,i}=0.5\,M_{\rm \odot}$, and $P_{\rm orb,i}=0.2\,\rm days$. Here $\dot J_{\rm MB}$ is the angular momentum loss rate of magnetic braking. The different colored lines correspond to different values of $f_{\rm dt}$: $f_{\rm dt}=0.03$ (blue), 0.05 (yellow), 0.1 (green), and 1.0 (cyan).}
 \label{Fig-compare-fdt}
\end{figure*}

\begin{figure*}[ht]
\centering
\plotone{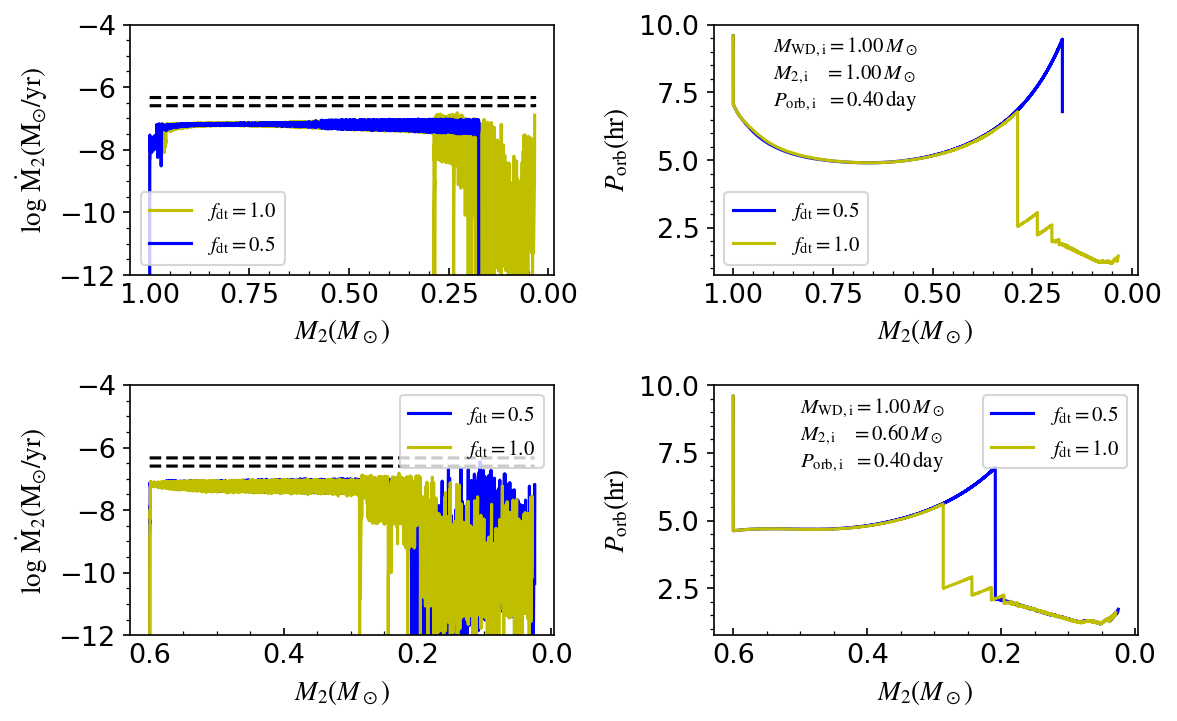}
 \caption{Similar to Fig.\,\ref{Fig-compare-fdt} but for $M_{\rm WD,i}=1.0\,M_{\rm \odot}$. In the upper and lower $M_{\rm 2,i}=1.0\,M_{\rm \odot}$, $P_{\rm orb,i}=0.4\,\rm days$, and $M_{\rm 2,i}=0.6\,M_{\rm \odot}$, $P_{\rm orb,i}=0.4\,\rm days$, respectively. The blue and yellow lines represent the results with $f_{\rm dt}=0.5$ and 1.0, respectively.}
 \label{Fig-compare-fdt1.0}
\end{figure*}

For systems with more massive WDs, the situation is more complicated. Fig.\,\ref{Fig-compare-fdt1.0} shows two examples with $M_{\rm WD}^{\rm i}=1.0\,M_{\rm \odot}$ and $P_{\rm orb}^{\rm i}=0.4\,\rm days$ but with different secondary masses, i.e. $M_{\rm 2}^{\rm i}=1.0\,M_{\rm \odot}$ and $M_{\rm 2}^{\rm i}=0.6\,M_{\rm \odot}$, respectively. We take $f_{\rm dt}$ to be 0.5 and 1.0. The value of $f_{\rm dt}$ cannot be too small for systems with massive WDs because the calculation will be very time-consuming due to the short recurrence time. For systems with $M_{2}^{\rm i}=0.6\,M_{\rm \odot}$ (lower panels), the mass transfer rates become different with different $f_{\rm dt}$ after MB ceases. For the evolution of orbital periods, the system with $f_{\rm dt}=1.0$ enters the period gap significantly earlier, while the orbital period with $f_{\rm dt}=0.5$ increases to 7.5\,hr before MB stops. The secondary can fill its Roche lobe again under gravitational radiation.
However, for the systems with $M_{2}^{\rm i}=1.0\,M_{\rm \odot}$, the orbit expands to $\sim 10$\,hr and the secondary cannot fill its Roche lobe again under the gravitational radiation with $f_{\rm dt}=0.5$. It should be emphasized that not all of systems with massive WDs finally become detached for $f_{\rm dt}=0.5$.

In summary, the smaller $dt$, the more precise the calculation should be. However, $dt$ cannot be too small because of time cost and the convergence of calculation. In order to investigate the overall properties of CVs, a large number of systems need to be calculated to obtain the parameter spaces that form CVs. So we have to make some compromises in the calculations. 

\end{appendix}

\end{document}